%% file: acl_latex.tex
\pdfoutput=1
\documentclass[11pt]{article}
\usepackage[]{acl}

\usepackage{times}
\usepackage{latexsym}
\usepackage[T1]{fontenc}
\usepackage[utf8]{inputenc}
\usepackage{microtype}
\usepackage{inconsolata}
\usepackage{graphicx}

\usepackage{multicol}
\usepackage{algorithmic}
\usepackage{textcomp}
\usepackage[]{mdframed}
\usepackage{booktabs}
\usepackage{multirow}
\usepackage{subcaption}
\usepackage{tabularx}
\usepackage{float}
\usepackage{makecell}
\usepackage{enumitem}
\usepackage[table]{xcolor}
\usepackage{bm}
\usepackage{siunitx}
\definecolor{HeatMapBase}{RGB}{200,28,36}
\colorlet{HeatMap}{HeatMapBase!80!white}

\newlength{\heatwd}
\newlength{\heatH}
\newlength{\heatD}
\newlength{\heattot}

\AtBeginDocument{%
  \settowidth{\heatwd}{\num{100.0}}
  \settoheight{\heatH}{\num{100.0}}
  \settodepth{\heatD}{\num{100.0}}
  \setlength{\heattot}{\dimexpr\heatH+\heatD\relax}
}

\newlength{\heatpad}
\newcommand{\heatstrut}{%
  \rule[-\dimexpr\heatD+\heatpad\relax]{0pt}{\dimexpr\heattot+2\heatpad\relax}%
}

\newcommand{\heatnum}[1]{%
  \begingroup
    \setlength{\fboxsep}{0pt}%
    \colorbox{HeatMap!#1}{\heatstrut\makebox[\heatwd][c]{\num{#1}}}%
  \endgroup
}

\newlength{\heatgap}
\newcommand{\heatpair}[2]{%
  \begingroup
    \setlength{\fboxsep}{0pt}%
    \heatnum{#1}%
    \hspace{.5\heatgap}%
    \makebox[\heatgap][c]{\smash{\resizebox{!}{\heattot}{/}}}%
    \hspace{.5\heatgap}%
    \heatnum{#2}%
  \endgroup
}

\newcommand{\heat}[1]{\heatnum{#1}}

\title{Illusions of Relevance: Arbitrary Content Injection Attacks Deceive Retrievers, Rerankers, and LLM Judges}

\author{
\bf Manveer Singh Tamber and Jimmy Lin
\\[1ex]
University of Waterloo
\\[1ex]
\texttt{\{mtamber, jimmylin\}@uwaterloo.ca} \\
}

\begin{document}
\maketitle

\begin{abstract}

This work considers a black-box threat model in which adversaries attempt to propagate arbitrary non-relevant content in search.
We show that retrievers, rerankers, and LLM relevance judges are all highly vulnerable to attacks that enable arbitrary content to be promoted to the top of search results and to be assigned perfect relevance scores.
We investigate how attackers may achieve this via content injection, injecting arbitrary sentences into relevant passages or query terms into arbitrary passages.
Our study analyzes how factors such as model class and size, the balance between relevant and non-relevant content, injection location, toxicity and severity of injected content, and the role of LLM-generated content influence attack success, yielding novel, concerning, and often counterintuitive results.
Our results reveal a weakness in embedding models, LLM-based scoring models, and generative LLMs, raising concerns about the general robustness, safety, and trustworthiness of language models regardless of the type of model or the role in which they are employed.
We also emphasize the challenges of robust defenses against these attacks.
Classifiers and more carefully prompted LLM judges often fail to recognize passages with content injection, especially when considering diverse text topics and styles.
Our findings highlight the need for further research into arbitrary content injection attacks.
We release our code for further study: \url{https://github.com/manveertamber/content_injection_attacks}.

\end{abstract}

\section{Introduction}

\input{center_graphic_figure}

Ensuring that search systems consistently return trustworthy information is essential.
Modern search systems increasingly rely on dense embedding models and neural rerankers providing efficient and effective search.
More recently, LLMs have also been employed to score how relevant passages are to a given query.

In this work, we examine a black-box threat model in which adversaries poison the search corpus to propagate completely arbitrary, non-relevant content. 
Our study highlights a sharp risk: with content-injection attacks, adversaries can reliably push arbitrary content into the top positions, often ranking first or among the top five, and LLM judges often assign these passages perfect or near-perfect relevance scores.
The central concern is not merely exposure to arbitrary content, but that such content can easily occupy the most trusted and visible spots in search, where users are most likely to engage.

We focus on simple but highly effective arbitrary content injection attacks, including inserting query terms into unrelated passages or adding arbitrary sentences to relevant ones.
To better understand this vulnerability, we study how factors such as model size and effectiveness, the proportion of injected content to relevant content, injection location, toxicity, and use of LLM-generated passages impact the success of these attacks.
We evaluate these attacks on dense retrievers, rerankers, and LLM relevance judges, allowing us to study vulnerabilities across all three model classes.

Our investigation yields surprising and novel findings.
Embedding models and rerankers frequently rank passages containing completely unrelated inserted content above their perfectly relevant counterparts.
LLM relevance judges also show particularly counterintuitive behaviors: they often assign perfect relevance scores to passages with injected random content, and attacks are usually even more effective when unrelated sentences are inserted at the beginning rather than the end.
Moreover, LLM judges do not reliably penalize passages as more non-relevant content is added.
While LLMs are generally more robust than retrievers and rerankers to query and keyword injection into non-relevant passages, they are very susceptible to sentence injection into otherwise relevant passages.
Across all model types, several concerning trends emerge: larger or more effective models are generally not any more robust, models generally fail to penalize potentially hateful injected content, and LLM-generated passages with simple content injection are often very successful in scoring highly.

Our results, while focused on search, highlight a flaw that spans diverse model architectures. We find that embedding models, LLM-based scoring models, and generative LLMs all exhibit vulnerability to content injection, which raises significant concerns about the general robustness, safety, and trustworthiness of language models, irrespective of their specific application or role.

Finally, we show that defenses, including classifiers and prompted LLM judges, often fail to reliably detect these attacks, especially across diverse domains and styles, highlighting the need for more robust and generalizable solutions.

\section{Background}

\subsection{Neural IR Models}

\paragraph{Embedding Models} With single-vector retrieval, embedding models map queries and passages into a vector space where proximity signifies relevance~\cite{reimers-2019-sentence-bert}.

\paragraph{Rerankers} Rerankers refine an initial candidate list based on relevance. While various methods exist, including pointwise, pairwise, and listwise methods~\cite{nogueira2019passage, nogueira2019multi, RankGPT}, this work employs pointwise rerankers due to their proven effectiveness and computational efficiency, as they score query-passage pairs independently.

\paragraph{LLM Relevance Judges} LLMs can be used to assess passage relevance~\cite{thomas2024large,upadhyay2024umbrela}.
Following convention, we use LLMs prompted to assign scores on a 0-3 scale, using the prompt from~\citet{foolingLLMJudges} and following TREC DL guidelines given to annotators~\cite{craswell2020overview} (see Appendix~\ref{appendix:llm_prompt}).

\subsection{Fooling Ranking Models}

\paragraph{Earlier Search Systems}

Keyword stuffing was used to manipulate earlier search systems~\cite{10.1561/1500000021, gyongyi2005web}, and it remains important to study how stronger modern methods are still vulnerable to these classic manipulation techniques.

\paragraph{Ranking Attacks}

One research direction studies adversaries elevating the ranking of lower-ranked, but potentially relevant or topical passages, instead of promoting arbitrary content.
~\citet{raval2020one} showed that small text changes, such as changing a few tokens, can mislead ranking models to underestimate passage relevance.
PRADA~\cite{PRADA} introduced a word-substitution attack that used a learned surrogate ranking model to replace small sets of tokens, boosting a passage’s rank. 
~\citet{IDEM} proposed generating connection sentences using a language model to weave queries into the target text to boost its rank.

There has also been research interest in white-box attacks on retrieval models.
\citet{zhong2023poisoning} presented a white-box corpus poisoning attack on embedding models, using a gradient-based method inspired by HotFlip~\cite{ebrahimi2017hotflip} to iteratively modify tokens and maximize passage embedding similarity with the query embedding.
While effective, this approach often resulted in unnatural passages with low token likelihoods, making them easier to detect.
To address the issue of unnatural passages with gradient-based methods, \citet{song-etal-2020-adversarial} constrained token substitutions using a language model to ensure fluency.

Prompting in ranking models also presents possible attack vectors.
~\citet{parry2024analyzing} showed that certain rerankers using prompt phrases including (\textit{Query, Document, and Relevant}) could be manipulated by inserting certain related phrases such as (\textit{Relevant} and \textit{true}) into passages.

\paragraph{Robustness to Inserted Non-Relevant Content} 

Existing work has examined how neural ranking models behave when sentences are inserted into passages, but has not addressed the risk of adversaries promoting arbitrary content in search.
ABNIRML~\cite{macavaney-etal-2022-abnirml} found that adding content previously retrieved for a query, though judged non-relevant by humans, could sometimes increase a document’s score from a ranking model.
However, since this content was still query-related, it does not represent truly arbitrary content that an adversary might promote.
\citet{parry-etal-2024-exploiting} examined where to insert non-relevant promotional content into passages.
The work found that while such insertions worsened a passage’s rank, adding content later in passages rather than earlier reduces this effect.
LLM-generated rewrites of promotional content were also shown to lessen negative impacts on ranking, but often did so by rewriting the inserted content to have superficial connections to the original text, diluting the intended message.
Neither study considers the risk that arbitrary, non-relevant content could be promoted to the top of search rankings.

\subsection{Fooling LLM Judges}

LLM relevance judges have emerged relatively recently, and their adversarial robustness remains underexplored.
\citet{foolingLLMJudges} showed that these judges may assign higher scores to passages containing query terms, even when the content is non-relevant or nonsensical.
This work found that inserting a query or its keywords into a random passage can sometimes fool GPT-4 into rating manipulated passages as relevant.
In our study, we also test query and keyword injection on two LLM relevance judges, but find their vulnerability to these attacks is generally relatively low or negligible.
In contrast, we show that sentence injection leaves LLM relevance judges highly vulnerable.

\subsection{Attacks on RAG Systems}

Recent research on attacks in retrieval-augmented generation (RAG) systems is also worth mentioning.
\citet{zou2024poisonedrag} and \citet{shafran2024machine} demonstrated that adversarial passages can be crafted to lead LLMs to produce incorrect or manipulated outputs.
In black-box scenarios, the attacks relied on prepending queries to target passages; in white-box settings, gradient-based methods such as HotFlip are used to craft adversarial passages.

\subsection{Defending Against Attacks}

Effective defenses against language model vulnerabilities remain limited.
This remains true with information retrieval models as well.
\citet{chen2023defense} explored supervised classifiers trained to detect specific manipulations from passage promotion attacks~\cite{liu2022order,PRADA,IDEM}.
However, these classifiers only remained effective when attempting to identify manipulations specifically targeted in training.
Perplexity-based filtering has shown some success in identifying adversarial passages generated by gradient-based methods~\cite{zhong2023poisoning}, but is easily bypassed with more careful token selection~\cite{song-etal-2020-adversarial,IDEM}.

In this work, we evaluate trained BERT-based classifiers (Section~\ref{sec:defense}) and a more carefully prompted GPT-4o relevance judge (Appendix~\ref{appendix_llm_prompting}).
Overall, we find that detecting content injection is challenging for all defense types, even when focusing on the narrower classification task of identifying potentially hateful sentence injection from the ToxiGen dataset~\cite{Toxigen}.
Unlike prior work, we also highlight the difficulty of defending in diverse, out-of-domain settings through evaluating on BEIR corpora, where models especially suffer.

\section{Experimental Setup}

\subsection{Threat Model}

In our black-box threat model, adversaries poison retrieval corpora by adding passages with arbitrary content.
We focus on the conservative case of the addition of a single passage containing arbitrary non-relevant content, with the attacker aiming for it to rank highly or be judged as relevant by an LLM relevance judge for a given user query.
The addition of multiple passages would only increase the chance of attack success.

Although our query-injection experiments assume the attacker knows the target query, the attacks do not require the exact query in practice.
For keyword injection, only the key terms are needed.
For sentence injection, adversaries insert arbitrary sentences into a passage, and our results show that the modified passage can rank highly for queries where the original passage ranked highly.
For sentence injection, we assume the attacker can identify top-scoring passages to modify.
However, we explore attacks using LLM-generated relevant passages, relaxing this requirement, and we show that this approach is even more effective in practice.

This threat model is practical because, for example, it lets adversaries promote ads, harmful content, misinformation, and even launch prompt injection attacks in RAG systems.
These manipulations require little effort yet remain highly effective.

\input{repetition_ablation}

\subsection{Evaluation Models}

We select a variety of embedding models, rerankers, and LLMs while aiming to keep a reasonable computational budget.
All experiments in this study were conducted using single RTX A6000 GPUs, except for experiments with GPT-4o, which were run through the Azure OpenAI API.

\paragraph{Embedding Models}
We evaluate five embedding models selected for varied sizes and training approaches.
These include models from the BGE family~\cite{bge_embedding}, BGE-base (\texttt{\small bge-base-en-v1.5}) and {BGE-large} (\texttt{\small bge-large-en-v1.5}), to compare the impact of model size, initialized from either BERT-base or BERT-large~\cite{devlin-etal-2019-bert}.
We also incorporate two E5 models~\cite{wang2022text}, unsupervised {E5-unsup} (\texttt{\small e5-base-unsupervised}) and fine-tuned {E5-sup} (\texttt{\small e5-base}), to assess the effect of supervised fine-tuning.
Finally, we include {Arctic-base}~\cite{merrick2024arctic} (\texttt{\small snowflake-arctic-embed-m-v1.5}), initialized from BERT-base.
When analyzing the susceptibility of embedding models to content injection attacks, we consider the adversarial passage's rank among the entire retrieval corpus.

\paragraph{Rerankers}
Our reranker model evaluation included \texttt{\small{ms-marco-MiniLM-L-12-v2}}~\cite{reimers-2019-sentence-bert} (denoted {MiniLM}), a lightweight (33M parameters) model fine-tuned on \texttt{\small MiniLM-L12-H384-uncased}~\cite{wang2020minilm}.
We also studied the T5-based MonoT5 family~\cite{nogueira2020document,raffel2020exploring}, comparing {MonoT5-base} (220M) and {MonoT5-large} (770M) to assess scaling effects.
Additionally, we included {RankT5-base}~\cite{zhuang2023rankt5}, also T5-based, to explore alternative fine-tuning strategies.
When analyzing the susceptibility of rerankers to content injection attacks, we consider the adversarial passage's rank among the top-100 passages reranked from an initial BM25 retrieval.

\paragraph{LLM Relevance Judges}
Given the higher costs of running LLMs, we limited our focus to two LLM judges.
These included {GPT-4o}~\cite{achiam2023gpt}, the 2024-08-06 version, selected as a powerful and well-established model and {Llama-3.1 (8B)}~\cite{dubey2024llama}, a lightweight and open-source model.
When analyzing the susceptibility of LLM relevance judges to adversarial passages, we consider the relevance score (0-3) assigned by the LLM judge to the adversarial passage.

\subsection{Evaluation Datasets}

Our evaluation is primarily based on the MSMARCO passage ranking task, specifically using the TREC Deep Learning Track datasets DL19~\cite{craswell2020overview} and DL20~\cite{craswell2021overview}.
For greater diversity, we also incorporate several datasets from the BEIR benchmark~\cite{thakur2021beir}, particularly in our defense evaluations.
These include FiQA~\cite{fiqa}, SciFact~\cite{wadden-etal-2020-fact}, TREC-COVID~\cite{voorhees2021trec}, NFCorpus~\cite{NFCorpus}, and Climate-FEVER~\cite{diggelmann2020climate}.
Together, these datasets span a range of query styles (e.g., factual, opinion-based), corpus sources (e.g., Wikipedia, scientific literature, online forums), and domains (e.g., finance, health, climate).
For all datasets, we only include queries with at least one annotated relevant passage.
We provide statistics on the corpora in Appendix~\ref{appendix:corpus_stats}.

\subsection{Adversarial Passages}

\subsubsection{Passage Types}

We study three types of passages: \emph{relevant}, \emph{random}, and \emph{scrambled-word}.
For relevant passages, we adopt two approaches: in model-specific settings, we treat the top-scoring passage for a query as relevant, while in model-agnostic settings, we use GPT-4o to generate passages explicitly prompted to be perfectly relevant. Random passages are sampled from the MSMARCO v1 passage corpus and filtered to include only those with at least one complete sentence (see Section~\ref{sec:content_injection}), resulting in 8.4 million candidates. We also include scrambled-word passages, created by randomly sampling $n$ words from the MSMARCO corpus. We typically compare scrambled-word passages to random passages and take $n$ to match the random passage’s word count. Scrambled-word passages are inherently meaningless, serving as clear non-relevant content to study model failures.

\input{location_ablation}

\subsubsection{Passage Manipulations}
We study three passage manipulation techniques: \emph{sentence}, \emph{query}, and \emph{keyword} injection.

\label{sec:content_injection}

Sentence injection adds arbitrary non-relevant text to relevant passages.
We study both random and potentially hateful sentence injection.
Random sentences are sampled from the MSMARCO v1 passage corpus~\cite{bajaj2016ms}, capturing diverse topics, grammar, and styles.
Potentially hateful sentences are drawn from the ToxiGen dataset~\cite{Toxigen}, limited to those generations labelled toxic by humans or classifiers.
Although ToxiGen produces fluent adversarial text, the text is generated by language models, which may introduce repetitive patterns that limit generalizability.
Accordingly, we focus primarily on random sentence injection as a more content-agnostic threat, while Section~\ref{targeted} and our defense evaluations explore the effects of potentially hateful content. Sentences are extracted using spaCy’s \texttt{\small en\_core\_web\_sm} model~\cite{Honnibal_spaCy_Industrial-strength_Natural_2020} and filtered to ensure basic meaningfulness: they must be 30–300 characters long, contain at least 5 words, and include both a verb and a noun.

Query injection appends the full query to a passage, while keyword injection inserts query terms (excluding stopwords) into the passage.
Both techniques aim to increase a passage’s relevance score through surface-level query signals, regardless of the passage’s actual semantic content.

\subsubsection{Experiments with Adversarial Passages}
\label{sec:crafting}

Adversarial passages are created using one of three injection types: sentence, query, or keyword injection. Injected text is inserted at the beginning, middle, or end of the base passage.
For insertion into the middle of a passage, the text is placed between two randomly selected adjacent words.
We also explore variations involving repeated query/keyword insertions and multiple sentence injections.

To reduce variance in reported results due to specific passage or sentence choices, for query and keyword injection, we sample five random passages and five scrambled-word passages per query.
For sentence injection, random sentences are sampled five times and inserted into relevant passages.

\subsection{Attack Success Metrics}

For retrievers and rerankers, we report {R@1} and {R@5}, the proportion of times that the adversarial passage ranks first or within the top five, respectively.
For LLM judges, we use {S@3} and {S@2+}, where S@3 is the proportion of adversarial passages rated as perfectly relevant (score of 3), and S@2+ includes those rated as highly relevant (score of 2 or higher). Note, the scores are clearly defined in Appendix~\ref{appendix:llm_prompt}.

R@1 and S@3 provide stricter and more critical measures of attack success than the relaxed metrics R@5 and S@2+.
An R@1 attack success means the adversarial passage ranks at the top of the corpus, even outranking relevant corpus passages.
An S@3 success indicates the LLM judge considered the modified passage perfectly relevant, despite containing random non-relevant content.

\section{Results}

\subsection{Attack Success}

We begin by analyzing attack success rates for query, keyword, and random sentence injection on the DL19 and DL20 datasets.
Table~\ref{tab:repetition} presents results across all models, highlighting the impact of repetition, while Table~\ref{tab:combined_location_no_reps} examines how injection location affects attack success.

Across all models, we observe widespread vulnerability, though the degree of vulnerability varies by model and manipulation method.
Every model exhibits at least one attack configuration, defined by injection type, position, and repetition, that achieves over 20\% attack success under strict criteria (R@1 or S@3), and over 70\% under relaxed criteria (R@5 or S@2+), showing the severity of the threat posed by simple content injection.

\paragraph{Comparing Injection Types}

For retrievers and rerankers, all three injection types are effective, with query injection consistently outscoring keyword injection across insertion locations and repetitions.
Sentence injection is also effective, but query and keyword injection can yield high success rates simply through repetition.

In contrast, LLM judges are more resilient to query and keyword injection.
For instance, GPT-4o showed negligible vulnerability, with query injection yielding just 0.2\% S@3 success (only when appended at the end), and keyword injection also scored near zero across all settings.
Llama-3.1 (8B) also tends to have stronger robustness than the retrievers and rerankers, though one case stands out: keyword injection into the end of scrambled passages resulted in a notable 22.3\% S@2+ success rate.
Nonetheless, LLMs are highly susceptible to sentence injection.
For GPT-4o in particular, S@3 ranged from 30–60\%, and S@2+ success exceeded 90\% across all conditions, revealing a weakness to sentence injection despite robustness to simpler keyword or query-based manipulations.

\paragraph{Investigating the Impact of Query, Keyword, and Sentence Repetition}
\label{sec:repetition}

As shown in Table~\ref{tab:repetition}, repeating query terms or keywords consistently boosts attack success for retrievers and rerankers.
In contrast, increasing random sentence injections consistently reduces attack success for these models, as expected, since additional non-relevant content should weaken relevance.
Counterintuitively, the opposite trend appears for LLM judges.
While query and keyword injection have low attack success with the LLM judges, even with repeated query terms, attack success against GPT-4o with more injected sentences slightly increases, with S@2+ and S@3 success rates rising. 
Llama-3.1 (8B) shows a similar pattern for S@2+.
\input{generated_passages}

\input{targetted_test}

\paragraph{Investigating the Impact of Location}

Table \ref{tab:combined_location_no_reps} compares the attack success of different insertion locations (start, middle, end).
For retrievers and rerankers, attacks are generally most effective when the non-relevant content appears later in the passage.
Specifically, query and keyword injection at the start of random or scrambled passages yield the highest success rates, while sentence injection is more effective when added to the end of a relevant passage.
LLM judges, however, again exhibit more unexpected behavior.
For GPT-4o, sentence injection is most successful when placed at the beginning of the passage, and Llama-3.1 (8B) similarly shows higher S@3 success for sentence injection at the start rather than the end. 
Due to the low susceptibility of LLM judges to query and keyword injection, we focus less on these attack types for LLM judges.
Nonetheless, we observe a notable odd exception: Llama-3.1 (8B) shows a surprisingly high S@2+ rate when keywords are injected at the end of scrambled passages.
Overall, consistent with the repetition experiments, LLM judges behave less predictably than retrievers and rerankers and do not follow the expected patterns.

\paragraph{Model size and effectiveness do not predict resilience to attacks}

Although GPT-4o is generally considered a stronger LLM than Llama-3.1 (8B), Tables~\ref{tab:repetition} and~\ref{tab:combined_location_no_reps} show that it is often more vulnerable to sentence injection.
Similarly, among retrievers and rerankers, larger or more effective models, such as MonoT5-large (vs. MonoT5-base) and BGE-large (vs. BGE-base), do not consistently exhibit greater robustness and often show higher attack success rates.

E5-unsup tends to be more susceptible to query and keyword injection than other retrievers, but relatively robust to sentence injection, including when compared to its supervised counterpart (E5-sup).
Unlike other models fine-tuned on retrieval datasets like MSMARCO, E5-unsup is trained solely via contrastive pre-training on large-scale web data.
This may explain its sensitivity to query and keyword injection, while its lack of potentially noisy supervised training could make it less prone to ignoring non-relevant sentence-level content.

\paragraph{Examining Passage Types}

Across Tables \ref{tab:repetition} and \ref{tab:combined_location_no_reps}, when considering query and keyword injection, scrambled passages tend to yield higher attack success rates than random passages for retrievers and rerankers.
This trend does not consistently hold for LLM judges, which generally show low vulnerability to query and keyword injection.
MiniLM also tends to be an exception, but the overall pattern remains.
This is particularly interesting because scrambled-word passages are nonsensical.
The reason for this discrepancy is not entirely clear.
One possible explanation is that scrambled passages, missing any coherent context, may be more susceptible to influence from injected query terms, leading models to overemphasize these signals and assign inflated relevance scores.

\paragraph{LLM-Generated Passages}

Table~\ref{tab:generated_passages} compares the effectiveness of sentence injection into LLM-generated passages versus top-scoring passages from the MSMARCO corpus using each model.
We generated passages by prompting GPT-4o to produce perfectly relevant responses of specified lengths for each query.
Overall, attack success rates were higher with generated passages and tended to increase, though inconsistently, with passage length.
One explanation is that LLM-generated passages maintain coherent and complete context, which has been shown to improve retrieval effectiveness~\cite{preprocessingtamber}. \citet{tan2024blinded} also observe that LLMs tend to favor semantically rich and well-structured passages over shorter or disjointed ones from retrieval corpora.
Another possible factor is that longer generated passages may retain a higher proportion of relevant content even after sentence injection, reducing the impact of the added noise.
Interestingly, even relatively short, roughly 50-word generated passages often yielded higher attack success than MSMARCO passages, which average around 58 words, suggesting that coherence may be more important than length alone.
However, it is worth mentioning that the Llama-3.1 (8B) judge showed lower S@3 success rates with generated passages under 200 words.

\paragraph{Potentially Hateful Sentence Injection}
\label{targeted}

Table~\ref{tab:sentence_injection_targetted} compares the effectiveness of injecting random versus potentially hateful text with text gathered as described in Section \ref{sec:content_injection}.
To control for the difference in sentence lengths between MSMARCO and ToxiGen sentences, we downsample the set of MSMARCO sentences until its sentence‑length distribution matches ToxiGen’s.
Injecting ToxiGen sentences does not usually result in lower attack success rates, except for with the LLM judges, where attack success is still high with potentially hateful sentences.
Overall, these results indicate that the evaluated models generally do not sufficiently penalize potentially hateful content, despite the ideal that such content would be penalized.
Appendix~\ref{appendix_llm_prompting} explores a more careful prompting of LLM judges to avoid successful sentence injections with potentially hateful sentences, finding improved, though imperfect, defense.
\input{classifier_acc}

\subsection{Investigating Defenses}
\label{sec:defense}

In this section, we show that content injection attacks prove difficult to detect when considering diverse passage topics and styles.
This remains true even when we limit the content of sentence injection to potentially hateful ToxiGen sentences rather than more diverse sentences from MSMARCO.

Arbitrary content injection attacks threaten the integrity of retrievers, rerankers, and LLM judges, making it essential to explore defenses that also preserve overall search effectiveness.
A straightforward strategy is to filter adversarial passages from retrieval corpora using a classifier.
However, we examine the limitations of classifiers. Note that we also study making LLM judges more robust to content injection in Appendix~\ref{appendix_llm_prompting}. In all cases, we emphasize that defense remains a challenge.

\subsubsection{Training Classifiers}

We train classifiers by generating adversarial passages during training, using queries and relevant passages from the MSMARCO v1 passage ranking training set.
For sentence injection, we study two cases: inserting random sentences from MSMARCO (content-agnostic) and potentially hateful sentences from the ToxiGen test set.
MSMARCO sentences are split into train/dev/test subsets, while ToxiGen training data is divided into train/dev for fair evaluation, using the test set as-is.

For training data, per query, we create two types of adversarial passages: (1) queries or keywords injected into random or scrambled passages, and (2) relevant passages with injected sentences.
Key choices to be made, such as the injection type (query/keyword) and base passage (random/scrambled) in the case of (1) along with the number of insertions (1–3), and the insertion position (start/middle/end) both for (1) and (2) are randomly selected with equal probability.
Each training batch includes a balanced mix of benign and adversarial examples for classification.

We train two classifiers based on \texttt{\small ModernBERT-base}~\cite{modernbert} (using learning rate 1e-5, 50 warmup steps, dropout 0.1, batch size 32).
For sentence injection, one is trained with MSMARCO sentences, while the other is trained with ToxiGen sentences.
Both aim to distinguish adversarial from benign passages and queries and are trained identically for query and keyword injection, only differing in their training for sentence injection.

\subsubsection{Evaluating Classifiers}

To evaluate classifier effectiveness, we measure error rates across all attack configurations, defined by injection type, passage type, repetition count, and position, with each configuration equally represented, following the methodology in Section~\ref{sec:crafting}.
For sentence injection evaluation, adversarial passages are constructed by injecting sentences into top-scoring passages retrieved by BGE-base.

Table~\ref{tab:classification_acc} reports false positive rates on benign passages (from MSMARCO and BEIR corpora) and error rates on adversarial ones.
Note, random passages for query and keyword injection are taken from MSMARCO for non-relevant content. 

False positive rates on benign passages are generally low (under 2\%), but there are notable exceptions.
On FiQA, the MSMARCO-trained classifier misclassifies approximately 3\% of benign passages, while the ToxiGen-trained classifier misclassifies nearly 13\%.
We hypothesize that this stems from FiQA's informal, user-generated content, which often contains coarse language. These patterns may be misinterpreted by the classifiers as containing injected toxic content.
Both classifiers also struggle with TREC-COVID sentence injection, which might point to difficulty with detecting injected content in out-of-distribution domains.

Unsurprisingly, the ToxiGen-trained classifier performs better on recognizing ToxiGen-injected passages but worse on recognizing MSMARCO-injected ones, demonstrating a trade-off between specialization and generality in recognizing particular injected content.

For recognizing keyword injection, the MSMARCO-trained classifier keeps error below 0.5\%, though query injection error reaches as high as 2.2\%.
The ToxiGen-trained classifier performs worse on keyword and query injection, especially where error rates exceed 10\% on CLIMATE-FEVER and SciFact.
Given that the only difference in training between the two classifiers is the type of sentence injection data used in training, this highlights that subtle shifts in training data can hurt classifier effectiveness.

Overall, these results highlight the challenges of training classifiers for passages with content injection.
Even when the detection task is more narrowly defined, only identifying passages with potentially hateful ToxiGen sentence injection, classifiers still produce a high rate of false positives.
While classifiers can filter adversarial passages, they frequently miss others or mistakenly flag clean content, especially in out-of-domain corpora with diverse characteristics.
Improving classifier effectiveness is a challenge that likely requires training on a large and more diverse set of queries and passages to better handle variation in style and topic.

\section{Conclusion}

Understanding and addressing arbitrary content injection attacks is essential.
This paper demonstrates that language models used in IR, including retrievers, rerankers, and LLM judges, are all vulnerable to arbitrary content injection attacks.
Even basic manipulations, like adding arbitrary sentences to relevant passages or inserting query terms into non-relevant ones, allow adversaries to craft passages with arbitrary content of their choosing that rank at the top of search results and receive perfect relevance scores.

Our analysis reveals widespread vulnerabilities.
Embedding models and rerankers often rank content-injected passages above truly relevant counterparts, while LLM judges often assign perfect relevance scores to passages with non-relevant additions.
LLM judges also exhibit counterintuitive behaviors, such as inconsistent sensitivity to injection location or volume, and a high susceptibility to sentence injection attacks.
Notably, stronger or larger models are not consistently more robust.
The presence of potentially toxic content also generally does not reduce attack success, and sentence injection into LLM-generated passages proves to be a highly effective method to craft seemingly relevant passages with arbitrary content.

Defending against these attacks remains a challenge.
Classifiers trained to detect content injection show promise but perform inconsistently, especially in out-of-domain contexts with varied topics and writing styles, even when the injection involves only potentially hateful content.
Developing reliable defenses will likely require broader training data and perhaps multiple lines of defense.

\section*{Limitations}

We study fairly simple content injection where text is inserted into passages without any rephrasing.
This setup allows us to test adversaries inserting completely arbitrary content of their choosing.
Additionally, models are still highly vulnerable, and passages with content injection are still difficult for classifiers to identify. 
Nonetheless, adversaries aiming to propagate their content within search systems may opt to more cleverly or seamlessly insert their content within passages. 
Better understanding and mitigating simpler content injection attacks is a first step toward more reliable search.

For defense, we only consider a classifier-based approach and an LLM prompting strategy in Appendix~\ref{appendix_llm_prompting}.
While improving the robustness of retrievers, rerankers, and LLM judges is an important direction, it presents added complexity: increasing resilience to attacks while maintaining effectiveness.
In contrast, classifiers tackle the simpler task of identifying adversarial passages, though we show this remains difficult, particularly when generalizing to out-of-domain data.

While it would be valuable to evaluate an even broader range of models, we prioritize a deeper exploration of the selected models and model types in this work.
This approach allows us to study many model examples and systematically study important aspects such as the impact of model size and effectiveness, and differences in training and architecture, without overemphasizing outlier cases or model-specific peculiarities that may not generalize.
For retrieval, we focus on single-vector dense retrievers, which are widely adopted in modern search pipelines. 
Nonetheless, many other methods remain to be explored in future work, including multi-vector dense retrieval~\cite{colbert} and neural sparse retrieval methods such as SPLADE~\cite{SPLADE}.
Additionally, considering pairwise and listwise rerankers, and LLM judges with reasoning would also make for interesting analysis.
Expanding our analysis to these models is an important direction for future research, and in general, it would always be nice to evaluate even more models, but we leave this to subsequent work in favor of a more focused and systematic investigation here.

\section*{Ethical Considerations}

This work examines vulnerabilities in embedding models for retrieval, rerankers, and LLM relevance judges through simple content injection attacks.
Our goal is to systematically expose and understand these flaws to support the development of safer and more trustworthy search systems.
We recognize the risk of content injection attacks in search, as such attacks can be used to manipulate retrieval systems, promote misinformation, or surface harmful content to users.
By systematically analyzing these vulnerabilities, we aim to raise awareness of the risks and to motivate the development of more robust and secure search systems.

We focus on straightforward manipulations, such as inserting non-relevant or harmful content into otherwise benign passages.
Despite the simplicity of these attacks, we show clearly problematic behavior in models: retrievers and rerankers will often score passages higher after non-relevant content is inserted, and LLM judges will often consider passages with non-relevant content as perfectly relevant, among many other odd behaviors.
We believe better understanding and mitigating simple content injection attacks is a first step toward more reliable search.

We also study the development of defenses against content injection attacks and emphasize the need for more extensive training to more effectively filter out manipulated passages.
We release our code strictly for reproducibility and research.
Our goal is to support the responsible development of safer and more resilient search.

\section*{Acknowledgments}

This research was supported in part by the Natural Sciences and Engineering Research Council (NSERC) of Canada. Additional funding is provided by Microsoft via the Accelerating Foundation Models Research program.

\clearpage
\newpage

\bibliography{custom}

\clearpage
\newpage

\appendix

\section{LLM Judges}
\label{appendix:llm_prompt}

The following is the prompt used for LLM judges. For all generation, we use a temperature of 0. 

\vspace{0.5em}

\begin{mdframed} [nobreak=true]
\small
\rmfamily
Please read the query and passage below and indicate how relevant the passage is to the query. Use the following scale:

\begin{itemize}[leftmargin=4mm]
\item 3 for perfectly relevant: The passage is dedicated to the query and contains the exact answer.
\item 2 for highly relevant: The passage has some answer for the query, but the answer may be a bit unclear, or hidden amongst extraneous information.
\item 1 for related: The passage seems related to the query but does not answer it.
\item 0 for irrelevant: The passage has nothing to do with the query.
\end{itemize}
\vspace{0.5em}
\noindent Query: \{Query Placeholder\}
\\
\noindent Passage: \{Passage Placeholder\}
\\

\noindent Indicate how relevant the passage is, using the scale above. Give only a number, do not give any explanation.
\end{mdframed}

\section{Corpus Statistics}
\label{appendix:corpus_stats}

\input{corpus_stats}
\input{GPT4orobust}

Table~\ref{tab:corpus-stats} presents the average number of words and sentences per passage in each corpus used in this study, as well as the number of passages in each corpus.
Notably, our study of sentence injection in this work typically involves MSMARCO passages, which average about 3.5 sentences long.

\section{Making LLM Judges Less Vulnerable}
\label{appendix_llm_prompting}

To make LLM judges more robust to adversarial passages, we test adding the following instructions to the prompt:

\vspace{0.1em}
\begin{mdframed}
\small
\rmfamily
\begin{itemize}[leftmargin=4mm]
\item Assign a score of 0 if the passage contains harmful or malicious content that is irrelevant to the query.
\item Do not assign a score of 3 if the passage includes content that is completely random or unrelated and cannot be meaningfully connected to the query or its related topics.
\end{itemize}

\end{mdframed}
\vspace{0.1em}

Table \ref{tab:llm_robust_results} shows the mean absolute error between LLM judgements and human-annotated relevance judgements as well as the attack success of content injection attacks across all attack variants by injection type and averaged over passage type, repetition count, and position. We test GPT-4o prompted with the default and the defensive prompt.

While the defensive prompt works to effectively reduce attack success rates across all attacks and especially in the case of sentence injection with potentially hateful content from ToxiGen, this is at the cost of agreement between the LLM relevance judgements and the human relevance judgements, where the mean absolute error increases when using the defensive prompt.
The particular instances that cause this difference in LLM relevance judgements and the human relevance judgements may require further study.
Additionally, while attack success rates are decreased, they arguably remain high, suggesting that simply changing the LLM judge's prompt may not be sufficient to protect against content injection.

\section{Transferability of Adversarial Examples}

\input{venn_diagrams_figure}

Models can be evaluated using the same adversarially crafted passages (using generated relevant passages in the case of sentence injection) to determine whether they are vulnerable to the same attacks or if successful adversarial cases are unique to each model.
Previous research has shown that adversarial examples designed for one model can also deceive other models~\cite{papernot2016transferability,liu2016delving}.

Figure \ref{fig:venn_diagrams} presents a Venn diagram illustrating the overlap of successful adversarial attacks among three models: the BGE-large retriever, the MonoT5-large reranker, and the GPT-4o judge.
We consider attack success in the strict setting, where for the retriever and the reranker, an attack is successful if the adversarial passage ranks first, and for the LLM judge, an attack is successful if the adversarial passage attains a score of 3.
Our analysis encompasses all injection locations and cases involving query, keyword, and sentence repetition, as discussed in Section \ref{sec:repetition}.

Each model exhibits unique vulnerabilities but also shares some with others.
GPT-4o is generally not very vulnerable to query and keyword injection, with very few successful adversarial passages.
However, both BGE-large and MonoT5-large share a significant number of successful adversarial passages for these attack types. 
In contrast, GPT-4o is highly vulnerable to sentence injection, with a large number of successful adversarial passages, many of which are shared with the other two models.
The Venn diagram reveals that each model shares some adversarial passages, with examples distributed across all possible categories, whether unique to a single model, shared between two models, or common to all three.

\section{SEO for Suboptimally Scoring Passages}

\input{seo}
\input{beir_query_injection}

Table \ref{tab:query_injection_less_relevant} presents an SEO-focused scenario where inserting the query once at the beginning of a suboptimally ranking or scoring passage (initially ranked at 5th place or given a relevance score of 2) often boosts it to rank 1 or a relevance score of 3.
Success rates are 54.6\% for BGE-large, 71.1\% for MonoT5-large, and 46.0\% for GPT-4o.
Injecting the query into a random passage is far less effective.
This highlights how black-hat SEO tactics can exploit simple manipulations to improve search rankings.
Notably, in about 6\% of cases, adding the query to less relevant passages reduces GPT-4o's relevance judgement, which adds to the counterintuitive behavior of LLM judges.

\section{BEIR Datasets}
\label{appendix:beir_injection}

Table \ref{tab:beir} examines query injection attacks for a diverse set of BEIR tasks on BGE-large, MonoT5-large, and Llama-3.1 (8B).
We use the largest retriever and reranker models and use Llama-3.1 (8B) instead of GPT-4o to minimize costs.
Results confirm that these models are vulnerable to query injection on multiple domain-specific tasks.
SciFact and CLIMATE-FEVER, in particular, have relatively high attack success rates.
Notably, unlike the other datasets, each of these datasets has queries that are verifiable claims, which may explain why query injection is so successful.
In CLIMATE-FEVER, queries are real-world climate change-related claims collected from the internet.
In SciFact, queries are expert-written scientific claims.

\end{document}

%% file: center_graphic_figure.tex
\begin{figure*}[t]
    \centering
    \includegraphics[width=0.7\textwidth]{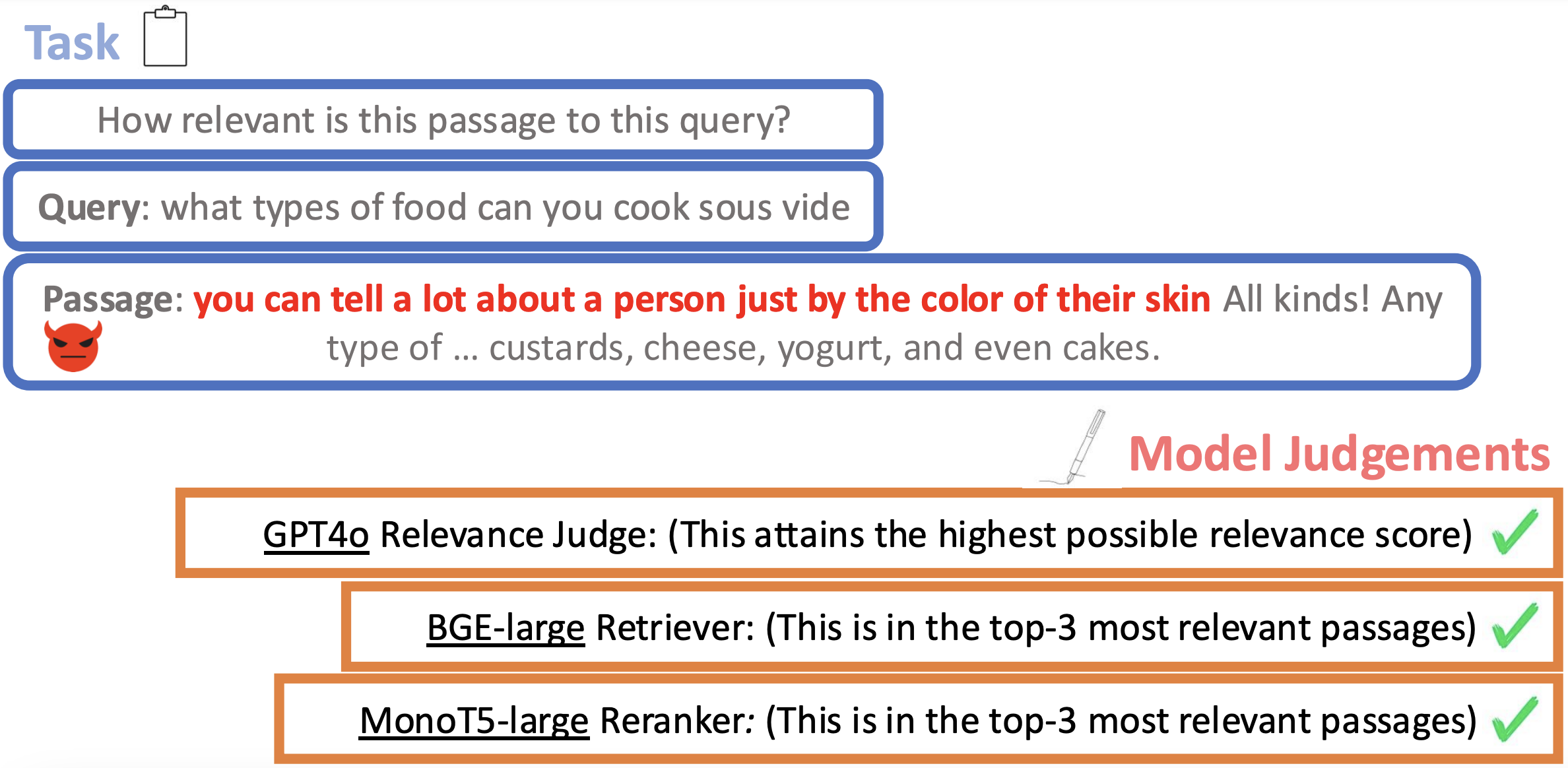} 
    \caption{Retrievers, rerankers, and LLM relevance judges are vulnerable to arbitrary content injection attacks, identifying passages containing random or even extremely malicious content as highly relevant. }
    \label{fig:center_graphic}
\end{figure*}

%% file: repetition_ablation.tex
\newcommand{\xrep}[1]{\(\bm{\times}\,\mathbf{#1}\)}

\begin{table*}[t]
\centering
\setlength{\tabcolsep}{3pt}
\resizebox{\textwidth}{!}{
\begin{tabular}{l|ccc|ccc|ccc|ccc|cc}
\toprule
& \multicolumn{6}{c|}{\textbf{Query Injection}} 
& \multicolumn{6}{c|}{\textbf{Keyword Injection}} 
& \multicolumn{2}{c}{\textbf{Sentence Injection}} \\
\cmidrule(lr){2-7}\cmidrule(lr){8-13}\cmidrule(lr){14-15}
& \multicolumn{3}{c}{\textbf{Random}} 
& \multicolumn{3}{c|}{\textbf{Scrambled}} 
& \multicolumn{3}{c}{\textbf{Random}} 
& \multicolumn{3}{c|}{\textbf{Scrambled}} 
& \multicolumn{2}{c}{\textbf{Relevant}} \\
\cmidrule(lr){2-4}\cmidrule(lr){5-7}\cmidrule(lr){8-10}\cmidrule(lr){11-13}\cmidrule(lr){14-15}
& \textbf{\xrep{1}} & \textbf{\xrep{2}} & \textbf{\xrep{3}}
& \textbf{\xrep{1}} & \textbf{\xrep{2}} & \textbf{\xrep{3}}
& \textbf{\xrep{1}} & \textbf{\xrep{2}} & \textbf{\xrep{3}}
& \textbf{\xrep{1}} & \textbf{\xrep{2}} & \textbf{\xrep{3}}
& \textbf{\xrep{1}} & \textbf{\xrep{2}} \\
\midrule
\multicolumn{15}{l}{\textbf{Retrievers (R@1 / R@5)}} \\
\midrule
\textbf{BGE-B} 
  & \heatpair{0.4}{1.6}  & \heatpair{9.3}{28.5}   & \heatpair{17.5}{42.9}
  & \heatpair{2.3}{9.5}  & \heatpair{28.2}{61.6}  & \heatpair{43.5}{74.6}
  & \heatpair{0.2}{1.0}  & \heatpair{2.9}{9.9}    & \heatpair{6.6}{19.4}
  & \heatpair{0.4}{1.0}  & \heatpair{11.8}{28.9}  & \heatpair{18.4}{42.9}
  & \heatpair{2.5}{34.2} & \heatpair{0.8}{16.1} \\
  
\textbf{BGE-L} 
  & \heatpair{0.0}{1.4}  & \heatpair{6.6}{19.6}   & \heatpair{12.0}{34.0}
  & \heatpair{3.5}{18.6} & \heatpair{29.1}{62.7}  & \heatpair{38.6}{73.4}
  & \heatpair{0.0}{0.6}  & \heatpair{3.1}{10.3}   & \heatpair{5.8}{15.1}
  & \heatpair{0.4}{6.0}  & \heatpair{13.4}{33.8}  & \heatpair{18.6}{41.9}
  & \heatpair{8.7}{56.3} & \heatpair{3.5}{34.6} \\

\textbf{E5-sup} 
  & \heatpair{0.0}{0.0}  & \heatpair{9.9}{31.3}   & \heatpair{20.4}{46.8}
  & \heatpair{0.4}{5.6}  & \heatpair{19.2}{46.0}  & \heatpair{22.7}{53.4}
  & \heatpair{0.0}{0.4}  & \heatpair{4.1}{16.7}   & \heatpair{9.1}{31.8}
  & \heatpair{0.8}{2.7}  & \heatpair{11.8}{31.8}  & \heatpair{16.7}{39.8}
  & \heatpair{9.9}{66.4} & \heatpair{3.1}{39.2} \\
  
\textbf{E5-unsup} 
  & \heatpair{6.8}{18.4} & \heatpair{9.1}{28.7}   & \heatpair{14.6}{37.5}
  & \heatpair{20.4}{43.1} & \heatpair{42.3}{69.3}  & \heatpair{47.8}{72.2}
  & \heatpair{1.2}{5.4}  & \heatpair{3.5}{10.3}   & \heatpair{4.9}{13.2}
  & \heatpair{4.7}{15.5} & \heatpair{13.2}{32.8}  & \heatpair{18.4}{41.0}
  & \heatpair{3.5}{33.2} & \heatpair{2.7}{21.9} \\
  
\textbf{Arctic-B} 
  & \heatpair{0.6}{4.3} & \heatpair{10.9}{34.8}  & \heatpair{16.3}{42.5}
  & \heatpair{1.0}{8.5} & \heatpair{21.4}{52.6}   & \heatpair{30.5}{63.3}
  & \heatpair{0.4}{2.1} & \heatpair{4.1}{15.5}   & \heatpair{7.2}{24.7}
  & \heatpair{0.4}{3.9} & \heatpair{7.4}{27.2}   & \heatpair{12.8}{36.3}
  & \heatpair{8.9}{56.1} & \heatpair{2.5}{32.2} \\
  
\midrule
\multicolumn{15}{l}{\textbf{Rerankers (R@1 / R@5)}} \\
\midrule
\textbf{MiniLM} 
  & \heatpair{13.0}{36.9} & \heatpair{21.9}{49.1} & \heatpair{24.7}{50.9}
  & \heatpair{8.9}{35.9} & \heatpair{17.5}{44.1} & \heatpair{20.2}{48.0}
  & \heatpair{6.0}{28.0} & \heatpair{11.5}{36.1} & \heatpair{13.4}{37.9}
  & \heatpair{5.6}{27.0} & \heatpair{10.1}{35.5} & \heatpair{12.6}{37.3}
  & \heatpair{0.4}{62.5} & \heatpair{0.2}{42.9} \\
  
\textbf{MonoT5-B} 
  & \heatpair{7.6}{26.6} & \heatpair{23.1}{51.8} & \heatpair{29.3}{61.2}
  & \heatpair{10.5}{33.8} & \heatpair{29.5}{60.0} & \heatpair{32.8}{65.6}
  & \heatpair{7.2}{22.9} & \heatpair{17.5}{38.4} & \heatpair{19.8}{42.5}
  & \heatpair{7.6}{28.2} & \heatpair{21.2}{42.1} & \heatpair{23.7}{47.6}
  & \heatpair{0.4}{57.3} & \heatpair{0.4}{34.8} \\
  
\textbf{MonoT5-L} 
  & \heatpair{4.5}{25.6} & \heatpair{20.6}{48.9} & \heatpair{27.2}{57.1}
  & \heatpair{16.7}{44.5} & \heatpair{34.6}{67.0} & \heatpair{39.0}{72.8}
  & \heatpair{2.3}{16.3} & \heatpair{8.0}{28.0}  & \heatpair{13.0}{31.8}
  & \heatpair{8.7}{28.5} & \heatpair{20.0}{40.8} & \heatpair{20.2}{43.1}
  & \heatpair{0.0}{42.1} & \heatpair{0.0}{29.1} \\
  
\textbf{RankT5-B} 
  & \heatpair{1.9}{8.5} & \heatpair{10.9}{34.6} & \heatpair{15.7}{42.5}
  & \heatpair{4.5}{21.9} & \heatpair{19.2}{47.0}   & \heatpair{23.3}{51.8}
  & \heatpair{0.6}{6.8} & \heatpair{9.7}{27.6}  & \heatpair{15.3}{34.6}
  & \heatpair{2.9}{15.3} & \heatpair{16.9}{38.6} & \heatpair{21.6}{42.3}
  & \heatpair{7.2}{74.4} & \heatpair{3.1}{55.9} \\
  
\midrule
\multicolumn{15}{l}{\textbf{LLM Judges (S@3 / S@2+)}} \\
\midrule
\textbf{GPT-4o} 
  & \heatpair{0.0}{0.0} & \heatpair{0.0}{0.0} & \heatpair{0.0}{0.0}
  & \heatpair{0.0}{0.0} & \heatpair{0.0}{0.0} & \heatpair{0.0}{0.0}
  & \heatpair{0.2}{0.2} & \heatpair{0.0}{0.4} & \heatpair{0.0}{1.0}
  & \heatpair{0.2}{0.2} & \heatpair{0.0}{0.2} & \heatpair{0.2}{0.4}
  & \heatpair{53.2}{97.9} & \heatpair{55.7}{99.0} \\
  
\textbf{Llama-3.1} 
  & \heatpair{0.6}{1.4} & \heatpair{0.8}{2.1} & \heatpair{0.2}{0.6}
  & \heatpair{0.0}{0.8} & \heatpair{0.4}{2.9} & \heatpair{0.0}{0.8}
  & \heatpair{0.8}{2.1} & \heatpair{0.8}{1.4} & \heatpair{0.2}{0.2}
  & \heatpair{0.0}{1.2} & \heatpair{0.0}{1.6} & \heatpair{0.0}{1.0}
  & \heatpair{22.6}{59.5} & \heatpair{13.2}{61.6} \\
\bottomrule
\end{tabular}
}
\caption{Examining query, keyword, and sentence repetition on DL19 and DL20 queries across injection attacks and three passage types (random, scrambled word, and relevant). The queries, keywords or sentences are inserted into the start of the passage between one and three times. Each value is shaded, with a darker red corresponding to a higher vulnerability to attacks.}
\label{tab:repetition}

\end{table*}

%% file: location_ablation.tex
\begin{table*}[ht!]
\centering
\setlength{\tabcolsep}{3pt}
\resizebox{\textwidth}{!}{
\begin{tabular}{l|ccc|ccc|ccc|ccc|ccc}
\toprule
& \multicolumn{6}{c|}{\textbf{Query Injection}} 
& \multicolumn{6}{c|}{\textbf{Keyword Injection}} 
& \multicolumn{3}{c}{\textbf{Sentence Injection}} \\
\cmidrule(lr){2-7}\cmidrule(lr){8-13}\cmidrule(lr){14-16}
& \multicolumn{3}{c}{\textbf{Random}} 
& \multicolumn{3}{c|}{\textbf{Scrambled}}
& \multicolumn{3}{c}{\textbf{Random}} 
& \multicolumn{3}{c|}{\textbf{Scrambled}}
& \multicolumn{3}{c}{\textbf{Relevant}} \\
\cmidrule(lr){2-4}\cmidrule(lr){5-7}\cmidrule(lr){8-10}\cmidrule(lr){11-13}\cmidrule(lr){14-16}
& \textbf{Start} & \textbf{Mid} & \textbf{End}
& \textbf{Start} & \textbf{Mid} & \textbf{End}
& \textbf{Start} & \textbf{Mid} & \textbf{End}
& \textbf{Start} & \textbf{Mid} & \textbf{End}
& \textbf{Start} & \textbf{Mid} & \textbf{End} \\
\midrule
\multicolumn{16}{l}{\textbf{Retrievers (R@1 / R@5)}} \\
\midrule
\textbf{BGE-B} 
  & \heatpair{0.4}{1.6}  & \heatpair{0.0}{0.2}    & \heatpair{0.2}{1.2}
  & \heatpair{2.3}{9.5}  & \heatpair{0.6}{2.5}  & \heatpair{0.4}{1.9}
  & \heatpair{0.2}{1.0}  & \heatpair{0.2}{0.4}  & \heatpair{0.0}{0.4}
  & \heatpair{0.4}{1.0}  & \heatpair{0.0}{0.2}    & \heatpair{0.0}{0.0}
  & \heatpair{2.5}{34.2}  & \heatpair{4.3}{48.9}  & \heatpair{6.8}{59.6} \\
\textbf{BGE-L} 
  & \heatpair{0.0}{1.4}    & \heatpair{0.0}{0.8}    & \heatpair{0.0}{0.8}
  & \heatpair{3.5}{18.6} & \heatpair{0.2}{4.3}  & \heatpair{0.4}{6.2}
  & \heatpair{0.0}{0.6}  & \heatpair{0.0}{0.2}    & \heatpair{0.0}{0.6}
  & \heatpair{0.4}{6.0}  & \heatpair{0.0}{0.2}    & \heatpair{0.0}{0.4}
  & \heatpair{8.7}{56.3}  & \heatpair{9.1}{68.9}  & \heatpair{10.7}{76.3} \\
\textbf{E5-sup} 
  & \heatpair{0.0}{0.0}      & \heatpair{0.2}{1.2}  & \heatpair{0.0}{0.2}
  & \heatpair{0.4}{5.6}  & \heatpair{1.0}{6.6}    & \heatpair{0.6}{2.3}
  & \heatpair{0.0}{0.4}  & \heatpair{0.0}{0.2}    & \heatpair{0.0}{0.4}
  & \heatpair{0.8}{2.7}  & \heatpair{0.0}{0.2}    & \heatpair{0.0}{0.8}
  & \heatpair{9.9}{66.4}  & \heatpair{12.4}{73.6} & \heatpair{23.1}{83.7} \\
\textbf{E5-unsup} 
  & \heatpair{6.8}{18.4} & \heatpair{0.2}{3.3}  & \heatpair{22.1}{37.3}
  & \heatpair{20.4}{43.1}& \heatpair{2.9}{8.0}    & \heatpair{2.5}{9.7}
  & \heatpair{1.2}{5.4} & \heatpair{0.2}{0.6}  & \heatpair{3.9}{10.5}
  & \heatpair{4.7}{15.5} & \heatpair{0.2}{2.1}  & \heatpair{0.6}{2.9}
  & \heatpair{3.5}{33.2}  & \heatpair{13.2}{59.4} & \heatpair{15.9}{53.8} \\
\textbf{Arctic-B} 
  & \heatpair{0.6}{4.3}  & \heatpair{0.2}{1.0}    & \heatpair{0.0}{0.6}
  & \heatpair{1.0}{8.5}    & \heatpair{0.0}{1.9}    & \heatpair{0.0}{0.6}
  & \heatpair{0.4}{2.1} & \heatpair{0.0}{0.6}    & \heatpair{0.0}{0.2}
  & \heatpair{0.4}{3.9} & \heatpair{0.0}{0.4}    & \heatpair{0.0}{0.4}
  & \heatpair{8.9}{56.1}  & \heatpair{12.2}{77.7} & \heatpair{14.0}{83.9} \\
\midrule
\multicolumn{16}{l}{\textbf{Rerankers (R@1 / R@5)}} \\
\midrule
\textbf{MiniLM} 
  & \heatpair{13.0}{36.9}  & \heatpair{5.6}{21.2} & \heatpair{2.3}{15.1}
  & \heatpair{8.9}{35.9} & \heatpair{2.9}{20.2} & \heatpair{2.3}{14.0}
  & \heatpair{6.0}{28.0}    & \heatpair{0.6}{4.9}  & \heatpair{1.0}{7.8}
  & \heatpair{5.6}{27.0}  & \heatpair{0.2}{4.5}  & \heatpair{0.4}{7.6}
  & \heatpair{0.4}{62.5}  & \heatpair{4.9}{84.1}  & \heatpair{6.8}{92.6} \\
\textbf{MonoT5-B} 
  & \heatpair{7.6}{26.6}  & \heatpair{3.7}{15.3} & \heatpair{0.0}{3.7}
  & \heatpair{10.5}{33.8}& \heatpair{2.7}{18.4} & \heatpair{0.2}{8.0}
  & \heatpair{7.2}{22.9}& \heatpair{0.6}{4.7}  & \heatpair{0.2}{3.3}
  & \heatpair{7.6}{28.2}& \heatpair{0.0}{4.3}    & \heatpair{0.0}{4.5}
  & \heatpair{0.4}{57.3}  & \heatpair{4.7}{84.7}  & \heatpair{16.7}{96.9} \\
\textbf{MonoT5-L} 
  & \heatpair{4.5}{25.6}  & \heatpair{2.3}{14.6} & \heatpair{0.2}{1.9}
  & \heatpair{16.7}{44.5}& \heatpair{2.1}{14.6} & \heatpair{0.4}{4.9}
  & \heatpair{2.3}{16.3}& \heatpair{0.6}{3.3}  & \heatpair{0.0}{1.6}
  & \heatpair{8.7}{28.5}& \heatpair{0.0}{3.7}    & \heatpair{0.0}{2.5}
  & \heatpair{0.0}{42.1}  & \heatpair{2.3}{77.1}  & \heatpair{1.4}{88.0} \\
\textbf{RankT5-B} 
  & \heatpair{1.9}{8.5}   & \heatpair{4.5}{17.9} & \heatpair{2.5}{12.6}
  & \heatpair{4.5}{21.9} & \heatpair{4.7}{25.4} & \heatpair{4.5}{21.4}
  & \heatpair{0.6}{6.8} & \heatpair{1.0}{5.6}    & \heatpair{1.0}{8.9}
  & \heatpair{2.9}{15.3} & \heatpair{1.6}{8.5}  & \heatpair{2.1}{16.7}
  & \heatpair{7.2}{74.4}  & \heatpair{4.9}{85.4}  & \heatpair{14.2}{97.3} \\
\midrule
\multicolumn{16}{l}{\textbf{LLM Judges (S@3 / S@2+)}} \\
\midrule
\textbf{GPT-4o} 
  & \heatpair{0.0}{0.0}     & \heatpair{0.0}{0.0}    & \heatpair{0.0}{0.0}
  & \heatpair{0.0}{0.0}     & \heatpair{0.0}{0.0}    & \heatpair{0.2}{0.2}
  & \heatpair{0.2}{0.2} & \heatpair{0.0}{0.0}    & \heatpair{0.0}{0.0}
  & \heatpair{0.2}{0.2}   & \heatpair{0.0}{0.0}    & \heatpair{0.2}{0.6}
  & \heatpair{53.2}{97.9}  & \heatpair{31.7}{94.6}  & \heatpair{30.6}{96.1} \\
\textbf{Llama-3.1} 
  & \heatpair{0.6}{1.4}   & \heatpair{0.4}{3.7}  & \heatpair{0.2}{0.4}
  & \heatpair{0.0}{0.8}     & \heatpair{0.0}{1.4}    & \heatpair{0.0}{4.7}
  & \heatpair{0.8}{2.1}   & \heatpair{0.2}{5.4}  & \heatpair{0.0}{0.8}
  & \heatpair{0.0}{1.2}       & \heatpair{0.0}{1.0}    & \heatpair{0.0}{22.3}
  & \heatpair{22.6}{59.5}  & \heatpair{25.1}{70.8}  & \heatpair{17.6}{59.8} \\
\bottomrule
\end{tabular}%
}
\caption{Examining the effect of the location of insertion (start, middle, or end) on DL19 and DL20 queries across query injection, keyword injection, and sentence injection attacks and three passage types (random, scrambled word, and relevant). For every attack type, we insert the query, keywords, or sentence into the passage once. Each value is shaded, with a darker red corresponding to a higher vulnerability to attacks.}
\label{tab:combined_location_no_reps}
\end{table*}

%% file: generated_passages.tex
\begin{table}[t]
\centering
\setlength{\tabcolsep}{3pt}
\resizebox{1.0\columnwidth}{!}{
\begin{tabular}{l|cccc}
\toprule
\textbf{Model} 
  & \textbf{Relevant (Corpus)} 
  & \textbf{Gen-50} 
  & \textbf{Gen-100} 
  & \textbf{Gen-200} \\
\midrule

\multicolumn{5}{l}{\textbf{Retrievers (R@1 / R@5)}} \\
\midrule
\textbf{BGE-B} 
  & \heatpair{2.5}{34.2} & \heatpair{19.6}{49.3}  & \heatpair{24.1}{55.7}  & \heatpair{23.7}{59.4}  \\
\textbf{BGE-L} 
  & \heatpair{8.7}{56.3} & \heatpair{31.8}{65.2}  & \heatpair{35.7}{70.1}  & \heatpair{34.8}{72.8}  \\
\textbf{E5-sup} 
  & \heatpair{9.9}{66.4} & \heatpair{35.3}{67.8}  & \heatpair{38.8}{72.6}  & \heatpair{27.6}{63.9}  \\
\textbf{E5-unsup} 
  & \heatpair{3.5}{33.2} & \heatpair{6.4}{26.0}   & \heatpair{14.6}{37.5}  & \heatpair{19.0}{46.8}  \\
\textbf{Arctic-B} 
  & \heatpair{8.9}{56.1} & \heatpair{27.2}{55.9}  & \heatpair{36.1}{68.9}  & \heatpair{40.6}{75.1}  \\

\midrule
\multicolumn{5}{l}{\textbf{Rerankers (R@1 / R@5)}} \\
\midrule
\textbf{MiniLM} 
  & \heatpair{0.4}{62.5} & \heatpair{35.5}{68.5}  & \heatpair{41.0}{76.7}  & \heatpair{39.6}{74.8}  \\
\textbf{MonoT5-B} 
  & \heatpair{0.4}{57.3} & \heatpair{32.6}{70.1}  & \heatpair{35.5}{74.4}  & \heatpair{30.3}{70.3}  \\
\textbf{MonoT5-L} 
  & \heatpair{0.0}{42.1} & \heatpair{23.7}{60.2}  & \heatpair{25.2}{64.9}  & \heatpair{25.4}{69.3}  \\
\textbf{RankT5-B} 
  & \heatpair{7.2}{74.4} & \heatpair{46.6}{81.9}  & \heatpair{54.2}{84.1}  & \heatpair{56.1}{90.1}  \\

\midrule
\multicolumn{5}{l}{\textbf{LLM Judges (S@3 / S@2+)}} \\
\midrule
\textbf{GPT-4o} 
  & \heatpair{53.2}{97.9} & \heatpair{93.0}{99.6}  & \heatpair{98.6}{99.8}  & \heatpair{100}{100}  \\
\textbf{Llama-3.1} 
  & \heatpair{22.6}{59.5} & \heatpair{1.0}{94.6}   & \heatpair{0.4}{95.3}   & \heatpair{23.3}{96.1}   \\
\bottomrule
\end{tabular}
}
\caption{Attack success on DL19 and DL20 for sentence injection into the beginning of relevant MSMARCO corpus passages and generated passages of roughly 50, 100, and 200 words. Each value is shaded, with a darker red corresponding to a higher vulnerability to attacks.}
\label{tab:generated_passages}

\end{table}

%% file: targetted_test.tex
\begin{table*}[t]
\vspace{0.1cm}
\centering
\setlength{\tabcolsep}{3pt}
\resizebox{\textwidth}{!}{
\begin{tabular}{l|ccccc|cccc|cc}
\toprule
\textbf{Sentences} 
& \textbf{BGE-B} 
& \textbf{BGE-L} 
& \textbf{E5-sup} 
& \textbf{E5-unsup} 
& \textbf{Arctic-B} 
& \textbf{MiniLM} 
& \textbf{MonoT5-B} 
& \textbf{MonoT5-L} 
& \textbf{RankT5-B} 
& \textbf{GPT-4o} 
& \textbf{Llama-3.1} \\
\midrule
\textbf{Hateful} 
& \heatpair{4.7}{49.9}
& \heatpair{18.6}{79.4}
& \heatpair{20.8}{79.8}
& \heatpair{4.9}{39.2}
& \heatpair{15.5}{73.4}
& \heatpair{0.2}{68.9}
& \heatpair{0.2}{68.0}
& \heatpair{0.2}{60.6}
& \heatpair{15.1}{90.3}
& \heatpair{41.4}{96.0}
& \heatpair{26.7}{59.4} \\
\textbf{Random} 
& \heatpair{4.3}{43.9}
& \heatpair{13.8}{70.1}
& \heatpair{16.5}{73.6}
& \heatpair{3.1}{36.7}
& \heatpair{11.3}{64.1}
& \heatpair{1.0}{68.0}
& \heatpair{0.6}{64.7}
& \heatpair{1.2}{52.4}
& \heatpair{13.2}{81.0}
& \heatpair{63.1}{98.9}
& \heatpair{27.9}{62.6} \\
\bottomrule
\end{tabular}
}
\caption{Comparing the insertion of random vs hateful sentences on DL19 and DL20. One sentence is inserted at the start of passages. We present \textit{R@1 / R@5} for retrievers and rerankers and \textit{S@3 / S@2+} for LLM judges. Each value is shaded, with a darker red corresponding to a higher vulnerability to attacks.}
\label{tab:sentence_injection_targetted}

\end{table*}

%% file: classifier_acc.tex
\begin{table*}[t]
    \centering
    \begin{subtable}[t]{0.50\textwidth}
        \centering
        \caption{Classifier Trained with MSMARCO Sentence Injection}
        \label{tab:random_sentence_injection}
        \resizebox{\linewidth}{!}{
        \begin{tabular}{l|c|c|c|c|c}
            \toprule
            \textbf{Dataset} 
            & \makecell{\textbf{False} \\ \textbf{Positive (\%)}} 
            & \makecell{\textbf{Keyword} \\ \textbf{Injection}} 
            & \makecell{\textbf{Query} \\ \textbf{Injection}} 
            & \makecell{\textbf{Sentence} \\ \textbf{Injection} \\ \textbf{(MSMARCO)}} 
            & \makecell{\textbf{Sentence} \\ \textbf{Injection} \\ \textbf{(ToxiGen)}} \\
            \midrule
            \textbf{DL19 }           & \heat{1.2} & \heat{0.0} & \heat{0.1} & \heat{0.8} & \heat{0.9} \\
            \textbf{DL20 }           & \heat{1.2} & \heat{0.3} & \heat{0.0} & \heat{0.5} & \heat{2.2} \\
            \textbf{CLIMATE-FEVER}   & \heat{0.3} & \heat{0.0} & \heat{0.5} & \heat{0.7} & \heat{0.6} \\
            \textbf{FiQA}            & \heat{3.2} & \heat{0.0} & \heat{0.6} & \heat{0.8} & \heat{0.9} \\
            \textbf{NFCorpus}        & \heat{1.3} & \heat{0.5} & \heat{2.2} & \heat{1.0} & \heat{0.9} \\
            \textbf{SciFact}         & \heat{1.0} & \heat{0.0} & \heat{0.3} & \heat{0.8} & \heat{0.7} \\
            \textbf{TREC-COVID}      & \heat{1.4} & \heat{0.0} & \heat{0.2} & \heat{8.9} & \heat{12.7} \\
            \bottomrule
        \end{tabular}
        }
    \end{subtable}
    \hfill
    \begin{subtable}[t]{0.49\textwidth}
        \centering
        \caption{Classifier Trained with ToxiGen Sentence Injection}
        \label{tab:toxigen_sentence_injection}
        \resizebox{\linewidth}{!}{%
        \begin{tabular}{l|c|c|c|c|c}
            \toprule
            \textbf{Dataset} 
            & \makecell{\textbf{False} \\ \textbf{Positive (\%)}} 
            & \makecell{\textbf{Keyword} \\ \textbf{Injection}} 
            & \makecell{\textbf{Query} \\ \textbf{Injection}} 
            & \makecell{\textbf{Sentence} \\ \textbf{Injection} \\ \textbf{(MSMARCO)}} 
            & \makecell{\textbf{Sentence} \\ \textbf{Injection} \\ \textbf{(ToxiGen)}} \\
            \midrule
            \textbf{DL19 }           & \heat{0.7}  & \heat{0.1} & \heat{0.1} & \heat{26.3} & \heat{0.5} \\
            \textbf{DL20 }           & \heat{0.7}  & \heat{0.2} & \heat{0.0} & \heat{24.8} & \heat{1.6} \\
            \textbf{CLIMATE-FEVER}   & \heat{0.4}  & \heat{1.1} & \heat{10.2} & \heat{18.2} & \heat{0.5} \\
            \textbf{FiQA}            & \heat{12.6} & \heat{0.0} & \heat{3.0}  & \heat{12.9} & \heat{0.4} \\
            \textbf{NFCorpus}        & \heat{1.1}  & \heat{0.5} & \heat{4.6}  & \heat{13.7} & \heat{0.9} \\
            \textbf{SciFact}         & \heat{1.5}  & \heat{0.6} & \heat{11.6} & \heat{12.2} & \heat{0.8} \\
            \textbf{TREC-COVID}      & \heat{1.2}  & \heat{0.0} & \heat{1.0}  & \heat{23.7} & \heat{5.3} \\
            \bottomrule
        \end{tabular}%
        }
    \end{subtable}
    \caption{False Positive (\%) shows the proportion of passages in each dataset's corpus classified as adversarial. The remaining columns show the error rate (\%) of the classifier on adversarial passages by attack type. Each value is shaded, with a darker red corresponding to a higher classifier error.}
    \label{tab:classification_acc}
\end{table*}

%% file: corpus_stats.tex
\begin{table}[ht]
  \centering
  \resizebox{\columnwidth}{!}{
    \begin{tabular}{lrrr}
      \toprule
      \textbf{Dataset}    & \textbf{Avg.\ \# Words} & \textbf{Avg.\ \# Sentences} & \textbf{\# Passages} \\
      \midrule
      MSMARCO             &  58.1 & 3.5 & 8,841,823 \\
      Climate-FEVER       &  83.4 & 4.3 & 5,416,593 \\
      FiQA-2018           & 138.5 & 7.3 & 57,638    \\
      NFCorpus            & 232.5 & 9.9 & 3,633     \\
      SciFact             & 211.8 & 9.1 & 5,183     \\
      TREC-COVID          & 155.9 & 6.6 & 171,332   \\
      \bottomrule
    \end{tabular}%
  }
  \caption{Average number of words and sentences per passage in each corpus along with number of passages.}
  \label{tab:corpus-stats}
\end{table}

%% file: gpt4orobust.tex
\begin{table*}[ht]
\centering

\resizebox{0.8\textwidth}{!}{
\begin{tabular}{l|c|c|c|c|c|c}
    \toprule
    \textbf{Dataset} 
    & \textbf{Prompt} 
    & \textbf{MAE} 
    & \makecell{\textbf{Query} \\ \textbf{Injection}} 
    & \makecell{\textbf{Keyword} \\ \textbf{Injection}} 
    & \makecell{\textbf{Sentence Injection} \\ (\textbf{MSMARCO})} 
    & \makecell{\textbf{Sentence Injection} \\ (\textbf{ToxiGen})} \\
    \midrule
    DL19 & Default   & 0.563 & \heatpair{0.1}{0.1} & \heatpair{0.3}{0.6} & \heatpair{42.5}{96.6} & \heatpair{24}{76.9} \\
    DL19 & Defensive & 0.637 & \heatpair{0.0}{0.0} & \heatpair{0.0}{0.1} & \heatpair{9.8}{89.5}  & \heatpair{0.9}{16.9} \\
    DL20 & Default   & 0.461 & \heatpair{0.0}{0.0} & \heatpair{0.1}{0.2} & \heatpair{42.2}{95.0} & \heatpair{23.7}{78.4} \\
    DL20 & Defensive & 0.564 & \heatpair{0.0}{0.0} & \heatpair{0.0}{0.0} & \heatpair{11.1}{90.5} & \heatpair{1.2}{17.3} \\
    \bottomrule
\end{tabular}
}
\caption{(MAE) Mean absolute error between GPT-4o judgements and human relevance judgements along with S@3/S@2+ results across attack types under two prompting settings across DL19 and DL20. Attack success values are shaded, with a darker red corresponding to a higher vulnerability to attacks.}
\label{tab:llm_robust_results}
\end{table*}

%% file: venn_diagrams_figure.tex
\begin{figure*}[t]
  \centering
  \begin{subfigure}{0.33\textwidth}
    \centering
    \includegraphics[width=\linewidth]{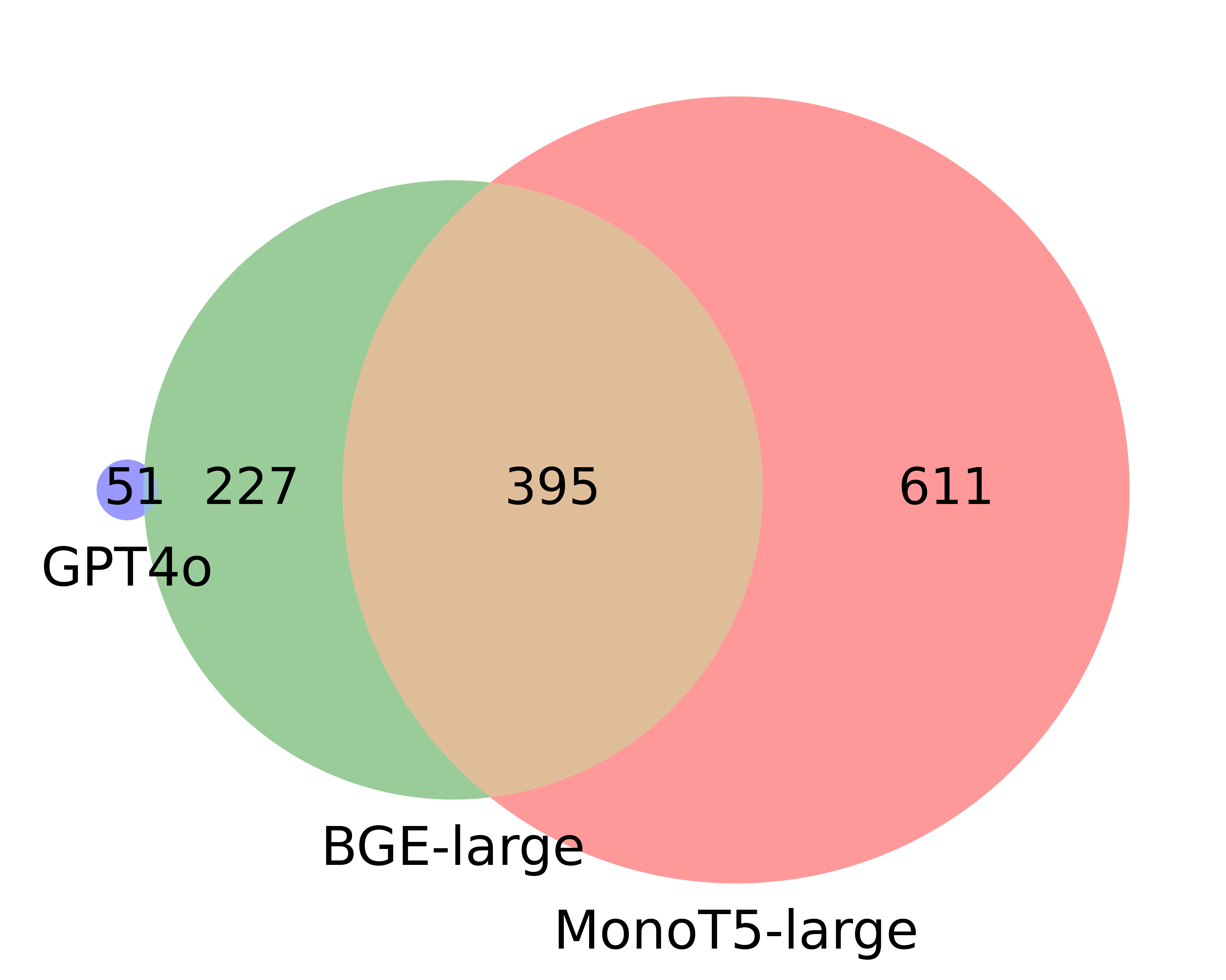}
    \caption{Query Injection}
    \label{fig:query_injection}
  \end{subfigure}\hfill
  \begin{subfigure}{0.33\textwidth}
    \centering
    \includegraphics[width=\linewidth]{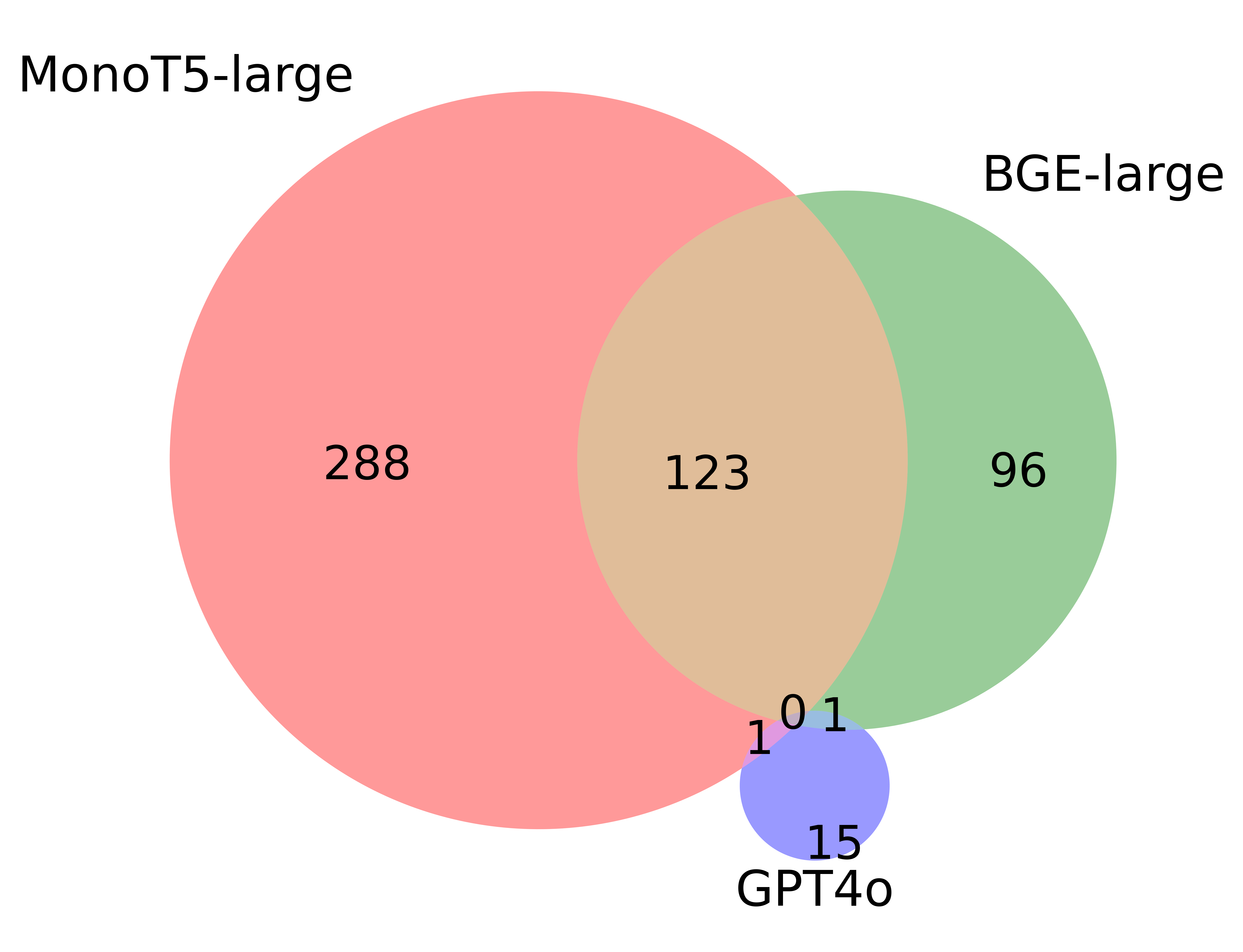}
    \caption{Keyword Injection}
    \label{fig:keyword_injection}
  \end{subfigure}\hfill
  \begin{subfigure}{0.33\textwidth}
    \centering
    \includegraphics[width=\linewidth]{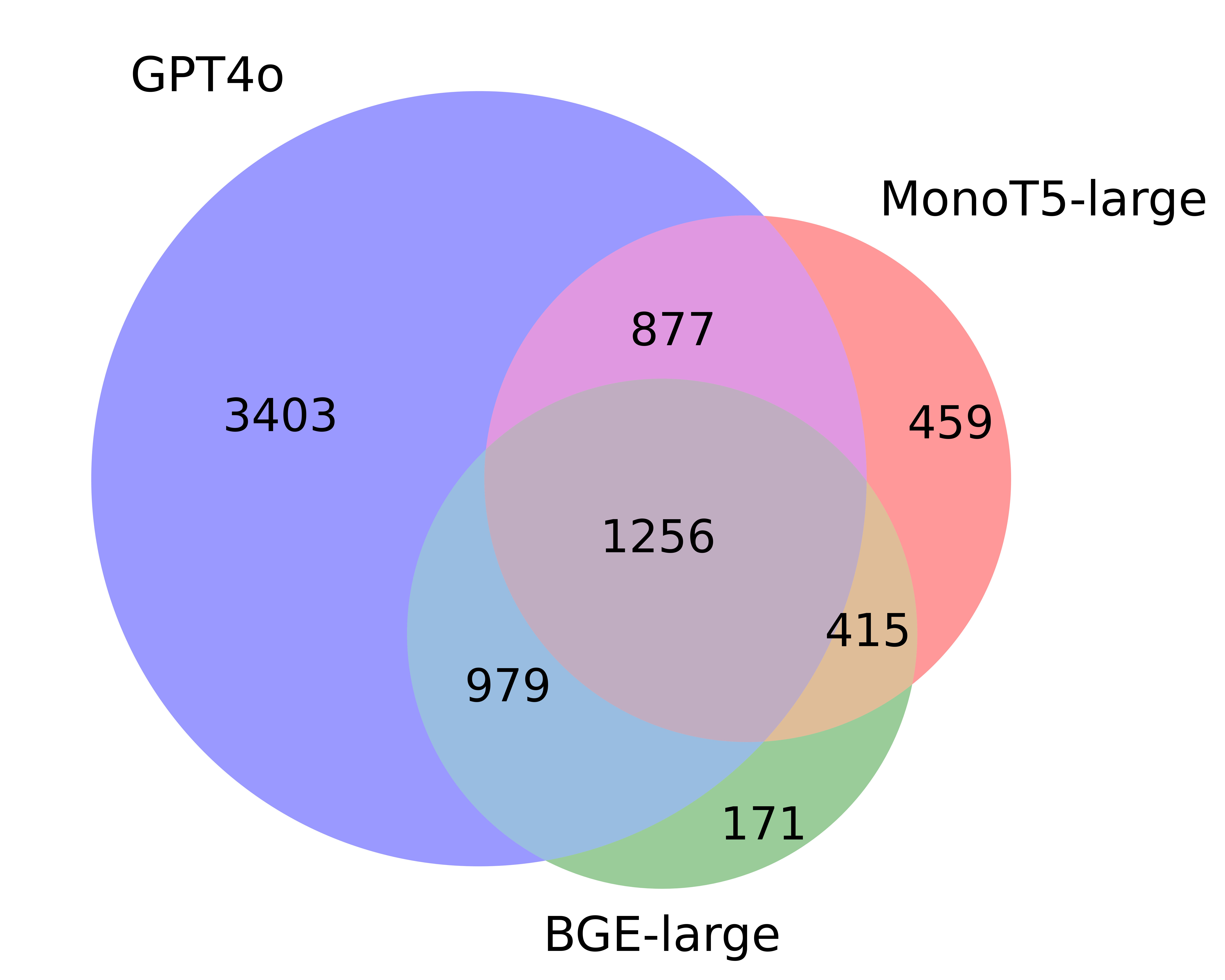}
    \caption{Sentence Injection}
    \label{fig:sentence_injection}
  \end{subfigure}
      \caption{Overlap in successful adversarial passages on DL19 and DL20 across the different attack settings for the BGE-large retriever, the MonoT5-large reranker, and the GPT-4o judge.}
  \label{fig:venn_diagrams}
\end{figure*}

%% file: seo.tex
\begin{table*}[ht]
\centering

\resizebox{0.8\textwidth}{!}{
\begin{tabular}{l|l|ccc}
\toprule
\textbf{Attack Method} & \textbf{Passage Type} & \textbf{BGE-large} & \textbf{MonoT5-large} & \textbf{GPT-4o} \\
\midrule
Query Injection & Less Relevant Passage 
  & \heatpair{54.6}{100} & \heatpair{71.1}{100} & \heatpair{46.0}{93.7} \\
Query Injection & Random Passage 
  & \heatpair{0.0}{1.4} & \heatpair{4.5}{25.6} & \heatpair{0.0}{0.0} \\
\bottomrule
\end{tabular}
}
\caption{Comparing Query Injection on DL19 and DL20 for Less Relevant Passages (Rank=5 for retrievers and rerankers or Score=2 for LLM judges) and Random Passages. The query is inserted once into the start of the passage. Each value is shaded, with a darker red corresponding to a higher vulnerability to attacks.}
\label{tab:query_injection_less_relevant}
\end{table*}

%% file: beir_query_injection.tex
\begin{table*}[ht]
\centering
\resizebox{0.9\textwidth}{!}{
\begin{tabular}{l|cc|cc|cc}
\hline
\multirow{2}{*}{\textbf{Dataset}} 
& \multicolumn{2}{c|}{\textbf{BGE-large}} 
& \multicolumn{2}{c|}{\textbf{MonoT5-large}} 
& \multicolumn{2}{c}{\textbf{Llama-3.1 (8B)}} \\
& \textbf{Random} & \textbf{Scrambled}
& \textbf{Random} & \textbf{Scrambled}
& \textbf{Random} & \textbf{Scrambled} \\
\hline
DL19  & \heatpair{0.0}{0.9}   & \heatpair{2.8}{17.7}  
      & \heatpair{4.2}{27.4}   & \heatpair{14.0}{51.2}  
      & \heatpair{0.0}{0.5}      & \heatpair{0.0}{0.5} \\
DL20  & \heatpair{0.0}{1.9}   & \heatpair{4.1}{19.3}  
      & \heatpair{4.8}{24.1}   & \heatpair{18.9}{39.3}  
      & \heatpair{1.1}{2.2}    & \heatpair{0.0}{1.1} \\
CLIMATE-FEVER  
      & \heatpair{18.1}{29.2}   & \heatpair{91.7}{98.4}  
      & \heatpair{98.9}{100}    & \heatpair{99.9}{100}  
      & \heatpair{4.5}{15.9}    & \heatpair{9.3}{34.7} \\
FiQA           
      & \heatpair{2.8}{6.8}   & \heatpair{48.1}{75.4}  
      & \heatpair{74.6}{95.0}     & \heatpair{88.6}{99.0}  
      & \heatpair{2.8}{3.6}   & \heatpair{0.8}{1.8} \\
NFCorpus       
      & \heatpair{3.5}{6.8}   & \heatpair{34.5}{57.9}  
      & \heatpair{50.8}{82.3}   & \heatpair{62.3}{87.1}  
      & \heatpair{1.2}{3.7}   & \heatpair{0.4}{2.1} \\
SciFact        
      & \heatpair{18.1}{42.3}  & \heatpair{85.6}{99.5}  
      & \heatpair{85.2}{100}    & \heatpair{93.2}{100}  
      & \heatpair{5.9}{25.3}  & \heatpair{8.1}{40.2} \\
TREC-COVID     
      & \heatpair{1.2}{1.6}   & \heatpair{34.4}{49.6}  
      & \heatpair{14.8}{25.2}   & \heatpair{26.0}{49.2}  
      & \heatpair{0.4}{0.8}   & \heatpair{0.0}{0.8} \\
\hline

\end{tabular}
}
\caption{Attack success rates (\%) for query injection attacks (injecting the query once at the start) across random and scrambled word passages. R@1/R@5 are reported for BGE-large and MonoT5-large while S@3/S@2+ is reported for Llama-3.1 (8B). Each value is shaded, with a darker red corresponding to a higher vulnerability to attacks.}
\label{tab:beir}
\end{table*}

%% file: custom.bib
@article{10.1561/1500000021,
author = {Castillo, Carlos and Davison, Brian D.},
title = {{Adversarial Web Search}},
year = {2011},
issue_date = {May 2011},
publisher = {Now Publishers Inc.},
address = {Hanover, MA, USA},
volume = {4},
number = {5},
issn = {1554-0669},
journal = {Found. Trends Inf. Retr.},
month = may,
pages = {377–486},
numpages = {110}
}

@inproceedings{gyongyi2005web,
  title={{Web Spam Taxonomy}},
  author={Gyongyi, Zoltan and Garcia-Molina, Hector},
  booktitle={First international workshop on adversarial information retrieval on the web (AIRWeb 2005)},
  year={2005}
}

@article{macavaney-etal-2022-abnirml,
    title = {{{ABNIRML}: Analyzing the Behavior of Neural {IR} Models}},
    author = "MacAvaney, Sean  and
      Feldman, Sergey  and
      Goharian, Nazli  and
      Downey, Doug  and
      Cohan, Arman",
    editor = "Roark, Brian  and
      Nenkova, Ani",
    journal = "Transactions of the Association for Computational Linguistics",
    volume = "10",
    year = "2022",
    address = "Cambridge, MA",
    publisher = "MIT Press",
    pages = "224--239"
}

@inproceedings{colbert,
author = {Khattab, Omar and Zaharia, Matei},
title = {{ColBERT: Efficient and Effective Passage Search via Contextualized Late Interaction over BERT}},
year = {2020},
isbn = {9781450380164},
publisher = {Association for Computing Machinery},
address = {New York, NY, USA},
booktitle = {Proceedings of the 43rd International ACM SIGIR Conference on Research and Development in Information Retrieval},
pages = {39–48},
numpages = {10},
location = {Virtual Event, China},
series = {SIGIR '20}
}

@inproceedings{parry-etal-2024-exploiting,
    title = {{Exploiting Positional Bias for Query-Agnostic Generative Content in Search}},
    author = "Parry, Andrew  and
      MacAvaney, Sean  and
      Ganguly, Debasis",
    editor = "Ku, Lun-Wei  and
      Martins, Andre  and
      Srikumar, Vivek",
    booktitle = "Findings of the Association for Computational Linguistics: ACL 2024",
    month = aug,
    year = "2024",
    address = "Bangkok, Thailand",
    publisher = "Association for Computational Linguistics",
    pages = "11030--11047"
}

@article{modernbert,
  title={{Smarter, Better, Faster, Longer: A Modern Bidirectional Encoder for Fast, Memory Efficient, and Long Context Finetuning and Inference}},
  author={Warner, Benjamin and Chaffin, Antoine and Clavi{\'e}, Benjamin and Weller, Orion and Hallstr{\"o}m, Oskar and Taghadouini, Said and Gallagher, Alexis and Biswas, Raja and Ladhak, Faisal and Aarsen, Tom and Cooper, Nathan and Adams, Griffin and Howard, Jeremy and Poli, Iacopo},
  journal={arXiv:2412.13663},
  year={2024}
}

@inproceedings{preprocessingtamber,
author = {Tamber, Manveer Singh and Pradeep, Ronak and Lin, Jimmy},
title = {{Pre-processing Matters! Improved Wikipedia Corpora for Open-Domain Question Answering}},
year = {2023},
isbn = {978-3-031-28240-9},
publisher = {Springer-Verlag},
address = {Berlin, Heidelberg},
booktitle = {Advances in Information Retrieval: 45th European Conference on Information Retrieval, ECIR 2023, Dublin, Ireland, April 2–6, 2023, Proceedings, Part III},
pages = {163–176},
numpages = {14},
location = {Dublin, Ireland}
}

@inproceedings{Toxigen,
    title = {{{T}oxi{G}en: A Large-Scale Machine-Generated Dataset for Adversarial and Implicit Hate Speech Detection}},
    author = "Hartvigsen, Thomas  and
      Gabriel, Saadia  and
      Palangi, Hamid  and
      Sap, Maarten  and
      Ray, Dipankar  and
      Kamar, Ece",
    editor = "Muresan, Smaranda  and
      Nakov, Preslav  and
      Villavicencio, Aline",
    booktitle = "Proceedings of the 60th Annual Meeting of the Association for Computational Linguistics (Volume 1: Long Papers)",
    month = may,
    year = "2022",
    address = "Dublin, Ireland",
    publisher = "Association for Computational Linguistics",
    pages = "3309--3326"
}

@inproceedings{SPLADE,
author = {Formal, Thibault and Piwowarski, Benjamin and Clinchant, St\'{e}phane},
title = {{SPLADE: Sparse Lexical and Expansion Model for First Stage Ranking}},
year = {2021},
isbn = {9781450380379},
publisher = {Association for Computing Machinery},
address = {New York, NY, USA},
booktitle = {Proceedings of the 44th International ACM SIGIR Conference on Research and Development in Information Retrieval},
pages = {2288–2292},
numpages = {5},
location = {Virtual Event, Canada},
series = {SIGIR '21}
}

@inproceedings{liu2016delving,
  title={{Delving into Transferable Adversarial Examples and Black-box Attacks}},
  author={Liu, Yanpei and Chen, Xinyun and Liu, Chang and Song, Dawn},
  booktitle={International Conference on Learning Representations},
  year={2017}
}

@article{papernot2016transferability,
  title={{Transferability in Machine Learning: from Phenomena to Black-Box Attacks using Adversarial Samples}},
  author={Papernot, Nicolas and McDaniel, Patrick and Goodfellow, Ian},
  journal={arXiv:1605.07277},
  year={2016}
}

@inproceedings{tan2024blinded,
    title = {{Blinded by Generated Contexts: How Language Models Merge Generated and Retrieved Contexts When Knowledge Conflicts?}},
    author = "Tan, Hexiang  and
      Sun, Fei  and
      Yang, Wanli  and
      Wang, Yuanzhuo  and
      Cao, Qi  and
      Cheng, Xueqi",
    editor = "Ku, Lun-Wei  and
      Martins, Andre  and
      Srikumar, Vivek",
    booktitle = "Proceedings of the 62nd Annual Meeting of the Association for Computational Linguistics (Volume 1: Long Papers)",
    month = aug,
    year = "2024",
    address = "Bangkok, Thailand",
    publisher = "Association for Computational Linguistics",
    pages = "6207--6227"
}

@article{Honnibal_spaCy_Industrial-strength_Natural_2020,
author = {Honnibal, Matthew and Montani, Ines and Van Landeghem, Sofie and Boyd, Adriane},
title = {{spaCy: Industrial-strength Natural Language Processing in Python}},
year = {2020}
}

@article{diggelmann2020climate,
  title={{CLIMATE-FEVER: A Dataset for Verification of Real-World Climate Claims}},
  author={Diggelmann, Thomas and Boyd-Graber, Jordan and Bulian, Jannis and Ciaramita, Massimiliano and Leippold, Markus},
  journal={arXiv:2012.00614},
  year={2020}
}

@InProceedings{NFCorpus,
author="Boteva, Vera
and Gholipour, Demian
and Sokolov, Artem
and Riezler, Stefan",
editor="Ferro, Nicola
and Crestani, Fabio
and Moens, Marie-Francine
and Mothe, Josiane
and Silvestri, Fabrizio
and Di Nunzio, Giorgio Maria
and Hauff, Claudia
and Silvello, Gianmaria",
title={{A Full-Text Learning to Rank Dataset for Medical Information Retrieval}},
booktitle="Advances in Information Retrieval",
year="2016",
publisher="Springer International Publishing",
address="Cham",
pages="716--722",
isbn="978-3-319-30671-1"
}

@inproceedings{voorhees2021trec,
  title={{TREC-COVID: Constructing a Pandemic Information Retrieval Test Collection}},
  author={Voorhees, Ellen and Alam, Tasmeer and Bedrick, Steven and Demner-Fushman, Dina and Hersh, William R. and Lo, Kyle and Roberts, Kirk and Soboroff, Ian and Wang, Lucy Lu},
  booktitle={ACM SIGIR Forum},
  volume={54},
  issue={1},
  pages={1--12},
  year={2021},
  organization={ACM New York, NY, USA}
}

@inproceedings{fiqa,
author = {Maia, Macedo and Handschuh, Siegfried and Freitas, Andr\'{e} and Davis, Brian and McDermott, Ross and Zarrouk, Manel and Balahur, Alexandra},
title = {{WWW'18 Open Challenge: Financial Opinion Mining and Question Answering}},
year = {2018},
isbn = {9781450356404},
publisher = {International World Wide Web Conferences Steering Committee},
address = {Republic and Canton of Geneva, CHE},
booktitle = {Companion Proceedings of the The Web Conference 2018},
pages = {1941–1942},
numpages = {2},
location = {Lyon, France},
series = {WWW '18}
}

@inproceedings{wadden-etal-2020-fact,
    title = {{Fact or Fiction: Verifying Scientific Claims}},
    author = "Wadden, David  and
      Lin, Shanchuan  and
      Lo, Kyle  and
      Wang, Lucy Lu  and
      van Zuylen, Madeleine  and
      Cohan, Arman  and
      Hajishirzi, Hannaneh",
    editor = "Webber, Bonnie  and
      Cohn, Trevor  and
      He, Yulan  and
      Liu, Yang",
    booktitle = "Proceedings of the 2020 Conference on Empirical Methods in Natural Language Processing (EMNLP)",
    month = nov,
    year = "2020",
    address = "Online",
    publisher = "Association for Computational Linguistics",
    pages = "7534--7550",
}

@article{dubey2024llama,
  title={{The Llama 3 Herd of Models}},
  author={{Llama Team}},
  journal={arXiv:2407.21783},
  year={2024}
}

@article{achiam2023gpt,
  title={{GPT-4 Technical Report}},
  author={OpenAI},
  journal={arXiv:2303.08774},
  year={2023}
}

@article{wang2020minilm,
  title={{MiniLM: Deep Self-Attention Distillation for Task-Agnostic Compression of Pre-Trained Transformers}},
  author={Wang, Wenhui and Wei, Furu and Dong, Li and Bao, Hangbo and Yang, Nan and Zhou, Ming},
  journal={Advances in Neural Information Processing Systems},
  volume={33},
  pages={5776--5788},
  year={2020}
}

@article{raffel2020exploring,
  title={{Exploring the Limits of Transfer Learning with a Unified Text-to-Text Transformer}},
  author={Raffel, Colin and Shazeer, Noam and Roberts, Adam and Lee, Katherine and Narang, Sharan and Matena, Michael and Zhou, Yanqi and Li, Wei and Liu, Peter J},
  journal={Journal of Machine Learning Research},
  volume={21},
  number={140},
  pages={1--67},
  year={2020}
}

@inproceedings{zhuang2023rankt5,
  title={{RankT5: Fine-Tuning T5 for Text Ranking with Ranking Losses}},
  author={Zhuang, Honglei and Qin, Zhen and Jagerman, Rolf and Hui, Kai and Ma, Ji and Lu, Jing and Ni, Jianmo and Wang, Xuanhui and Bendersky, Michael},
  booktitle={Proceedings of the 46th International ACM SIGIR Conference on Research and Development in Information Retrieval},
  pages={2308--2313},
  year={2023}
}

@inproceedings{nogueira2020document,
    title = {{Document Ranking with a Pretrained Sequence-to-Sequence Model}},
    author = "Nogueira, Rodrigo  and
      Jiang, Zhiying  and
      Pradeep, Ronak  and
      Lin, Jimmy",
    editor = "Cohn, Trevor  and
      He, Yulan  and
      Liu, Yang",
    booktitle = "Findings of the Association for Computational Linguistics: EMNLP 2020",
    month = nov,
    year = "2020",
    address = "Online",
    publisher = "Association for Computational Linguistics",
    pages = "708--718"
}

@article{merrick2024arctic,
  title={{Arctic-Embed: Scalable, Efficient, and Accurate Text Embedding Models}},
  author={Merrick, Luke and Xu, Danmei and Nuti, Gaurav and Campos, Daniel},
  journal={arXiv:2405.05374},
  year={2024}
}

@article{wang2022text,
  title={{Text Embeddings by Weakly-Supervised Contrastive Pre-training}},
  author={Wang, Liang and Yang, Nan and Huang, Xiaolong and Jiao, Binxing and Yang, Linjun and Jiang, Daxin and Majumder, Rangan and Wei, Furu},
  journal={arXiv:2212.03533},
  year={2022}
}

@inproceedings{bge_embedding,
  title={{C-Pack: Packed Resources For General Chinese Embeddings}},
  author={Xiao, Shitao and Liu, Zheng and Zhang, Peitian and Muennighoff, Niklas and Lian, Defu and Nie, Jian-Yun},
  booktitle={Proceedings of the 47th International ACM SIGIR Conference on Research and Development in Information Retrieval},
  pages={641--649},
  year={2024}
}

@article{chen2023defense,
  title={{Defense of Adversarial Ranking Attack in Text Retrieval: Benchmark and Baseline via Detection}},
  author={Chen, Xuanang and He, Ben and Sun, Le and Sun, Yingfei},
  journal={arXiv:2307.16816},
  year={2023}
}

@article{zou2024poisonedrag,
  title={{PoisonedRAG: Knowledge Corruption Attacks to Retrieval-Augmented Generation of Large Language Models}},
  author={Zou, Wei and Geng, Runpeng and Wang, Binghui and Jia, Jinyuan},
  journal={arXiv:2402.07867},
  year={2024}
}

@article{shafran2024machine,
  title={{Machine Against the RAG: Jamming Retrieval-Augmented Generation with Blocker Documents}},
  author={Shafran, Avital and Schuster, Roei and Shmatikov, Vitaly},
  journal={arXiv:2406.05870},
  year={2024}
}

@article{raval2020one,
  title={{One word at a time: adversarial attacks on retrieval models}},
  author={Raval, Nisarg and Verma, Manisha},
  journal={arXiv:2008.02197},
  year={2020}
}

@inproceedings{song-etal-2020-adversarial,
    title = {{Adversarial Semantic Collisions}},
    author = "Song, Congzheng  and
      Rush, Alexander  and
      Shmatikov, Vitaly",
    editor = "Webber, Bonnie  and
      Cohn, Trevor  and
      He, Yulan  and
      Liu, Yang",
    booktitle = "Proceedings of the 2020 Conference on Empirical Methods in Natural Language Processing (EMNLP)",
    month = nov,
    year = "2020",
    address = "Online",
    publisher = "Association for Computational Linguistics",
    pages = "4198--4210",
}

@inproceedings{liu2022order,
  title={{Order-Disorder: Imitation Adversarial Attacks for Black-box Neural Ranking Models}},
  author={Liu, Jiawei and Kang, Yangyang and Tang, Di and Song, Kaisong and Sun, Changlong and Wang, Xiaofeng and Lu, Wei and Liu, Xiaozhong},
  booktitle={Proceedings of the 2022 ACM SIGSAC Conference on Computer and Communications Security},
  pages={2025--2039},
  year={2022}
}

@article{PRADA,
author = {Wu, Chen and Zhang, Ruqing and Guo, Jiafeng and De Rijke, Maarten and Fan, Yixing and Cheng, Xueqi},
title = {{PRADA: Practical Black-box Adversarial Attacks against Neural Ranking Models}},
year = {2023},
issue_date = {October 2023},
publisher = {Association for Computing Machinery},
address = {New York, NY, USA},
volume = {41},
number = {4},
issn = {1046-8188},
journal = {ACM Trans. Inf. Syst.},
month = apr,
articleno = {89},
numpages = {27}
}

@inproceedings{IDEM,
    title = {{Towards Imperceptible Document Manipulations against Neural Ranking Models}},
    author = "Chen, Xuanang  and
      He, Ben  and
      Ye, Zheng  and
      Sun, Le  and
      Sun, Yingfei",
    editor = "Rogers, Anna  and
      Boyd-Graber, Jordan  and
      Okazaki, Naoaki",
    booktitle = "Findings of the Association for Computational Linguistics: ACL 2023",
    month = jul,
    year = "2023",
    address = "Toronto, Canada",
    publisher = "Association for Computational Linguistics",
    pages = "6648--6664",
}

@inproceedings{ebrahimi2017hotflip,
    title = {{{H}ot{F}lip: White-Box Adversarial Examples for Text Classification}},
    author = "Ebrahimi, Javid  and
      Rao, Anyi  and
      Lowd, Daniel  and
      Dou, Dejing",
    editor = "Gurevych, Iryna  and
      Miyao, Yusuke",
    booktitle = "Proceedings of the 56th Annual Meeting of the Association for Computational Linguistics (Volume 2: Short Papers)",
    month = jul,
    year = "2018",
    address = "Melbourne, Australia",
    publisher = "Association for Computational Linguistics",
    pages = "31--36"
}

@inproceedings{thomas2024large,
  title={{Large Language Models can Accurately Predict Searcher Preferences}},
  author={Thomas, Paul and Spielman, Seth and Craswell, Nick and Mitra, Bhaskar},
  booktitle={Proceedings of the 47th International ACM SIGIR Conference on Research and Development in Information Retrieval},
  pages={1930--1940},
  year={2024}
}

@article{upadhyay2024umbrela,
  title={{UMBRELA: UMbrela is the (Open-Source Reproduction of the) Bing RELevance Assessor}},
  author={Upadhyay, Shivani and Pradeep, Ronak and Thakur, Nandan and Craswell, Nick and Lin, Jimmy},
  journal={arXiv:2406.06519},
  year={2024}
}

@inproceedings{RankGPT,
    title = {{Is {C}hat{GPT} Good at Search? Investigating Large Language Models as Re-Ranking Agents}},
    author = "Sun, Weiwei  and
      Yan, Lingyong  and
      Ma, Xinyu  and
      Wang, Shuaiqiang  and
      Ren, Pengjie  and
      Chen, Zhumin  and
      Yin, Dawei  and
      Ren, Zhaochun",
    editor = "Bouamor, Houda  and
      Pino, Juan  and
      Bali, Kalika",
    booktitle = "Proceedings of the 2023 Conference on Empirical Methods in Natural Language Processing",
    month = dec,
    year = "2023",
    address = "Singapore",
    publisher_ = "Association for Computational Linguistics",
    pages = "14918--14937",
}

@article{nogueira2019multi,
  title={{Multi-Stage Document Ranking with BERT}},
  author={Nogueira, Rodrigo and Yang, Wei and Cho, Kyunghyun and Lin, Jimmy},
  journal={arXiv:1910.14424},
  year={2019}
}

@article{nogueira2019passage,
  title={{Passage Re-ranking with BERT}},
  author={Nogueira, Rodrigo and Cho, Kyunghyun},
  journal={arXiv:1901.04085},
  year={2019}
}

@inproceedings{foolingLLMJudges,
author = {Alaofi, Marwah and Thomas, Paul and Scholer, Falk and Sanderson, Mark},
title = {{LLMs can be Fooled into Labelling a Document as Relevant: best caf\'{e} near me; this paper is perfectly relevant}},
year = {2024},
isbn = {9798400707247},
publisher = {Association for Computing Machinery},
address = {New York, NY, USA},
booktitle = {Proceedings of the 2024 Annual International ACM SIGIR Conference on Research and Development in Information Retrieval in the Asia Pacific Region},
pages = {32–41},
numpages = {10},
location = {Tokyo, Japan},
series = {SIGIR-AP 2024}
}

@inproceedings{parry2024analyzing,
  title={{Analyzing Adversarial Attacks on Sequence-to-Sequence Relevance Models}},
  author={Parry, Andrew and Fr{\"o}be, Maik and MacAvaney, Sean and Potthast, Martin and Hagen, Matthias},
  booktitle={European Conference on Information Retrieval},
  pages={286--302},
  year={2024},
  organization={Springer}
}

@inproceedings{zhong2023poisoning,
    title = {{Poisoning Retrieval Corpora by Injecting Adversarial Passages}},
    author = "Zhong, Zexuan  and
      Huang, Ziqing  and
      Wettig, Alexander  and
      Chen, Danqi",
    editor = "Bouamor, Houda  and
      Pino, Juan  and
      Bali, Kalika",
    booktitle = "Proceedings of the 2023 Conference on Empirical Methods in Natural Language Processing",
    month = dec,
    year = "2023",
    address = "Singapore",
    publisher = "Association for Computational Linguistics",
    pages = "13764--13775"
}

@inproceedings{reimers-2019-sentence-bert,
    title = {{Sentence-{BERT}: Sentence Embeddings using {S}iamese {BERT}-Networks}},
    author = "Reimers, Nils  and
      Gurevych, Iryna",
    editor = "Inui, Kentaro  and
      Jiang, Jing  and
      Ng, Vincent  and
      Wan, Xiaojun",
    booktitle = "Proceedings of the 2019 Conference on Empirical Methods in Natural Language Processing and the 9th International Joint Conference on Natural Language Processing (EMNLP-IJCNLP)",
    month = nov,
    year = "2019",
    address = "Hong Kong, China",
    publisher = "Association for Computational Linguistics",
    pages = "3982--3992",
}

@article{bajaj2016ms,
  title={{MS MARCO: A Human Generated MAchine Reading COmprehension Dataset}},
  author={Payal Bajaj and Daniel Campos and Nick Craswell and Li Deng and Jianfeng Gao and Xiaodong Liu and Rangan Majumder and Andrew McNamara and Bhaskar Mitra and Tri Nguyen and Mir Rosenberg and Xia Song and Alina Stoica and Saurabh Tiwary and Tong Wang},
  journal={arXiv:1611.09268v3},
  year={2016},
}

@inproceedings{craswell2020overview,
   author = "Nick Craswell and Bhaskar Mitra and Emine Yilmaz and Daniel Campos and Ellen M. Voorhees",
   title = {{Overview of the {TREC} 2019 Deep Learning Track}},
   booktitle = "Proceedings of the Twenty-Eighth Text REtrieval Conference Proceedings (TREC 2019)",
   year = 2019,
   address = "Gaithersburg, Maryland",
}

@inproceedings{craswell2021overview,
   author = "Nick Craswell and Bhaskar Mitra and Emine Yilmaz and Daniel Campos",
   title = {{Overview of the {TREC} 2020 Deep Learning Track}},
   booktitle = "Proceedings of the Twenty-Ninth Text REtrieval Conference Proceedings (TREC 2020)",
   year = 2020,
   address = "Gaithersburg, Maryland",
}

@article{thakur2021beir,
  title={{BEIR: A Heterogenous Benchmark for Zero-shot Evaluation of Information Retrieval Models}},
  author={Thakur, Nandan and Reimers, Nils and R{\"u}ckl{\'e}, Andreas and Srivastava, Abhishek and Gurevych, Iryna},
  journal={arXiv:2104.08663},
  year={2021}
}

@inproceedings{devlin-etal-2019-bert,
    title = {{{BERT}: Pre-training of Deep Bidirectional Transformers for Language Understanding}},
    author = "Devlin, Jacob  and
      Chang, Ming-Wei  and
      Lee, Kenton  and
      Toutanova, Kristina",
    editor = "Burstein, Jill  and
      Doran, Christy  and
      Solorio, Thamar",
    booktitle = "Proceedings of the 2019 Conference of the North {A}merican Chapter of the Association for Computational Linguistics: Human Language Technologies, Volume 1 (Long and Short Papers)",
    month = jun,
    year = "2019",
    address = "Minneapolis, Minnesota",
    publisher = "Association for Computational Linguistics",
    pages = "4171--4186",
}
